\documentclass[11pt]{article}
\pdfoutput=1
\usepackage{jheppub}
\usepackage{hyperref}
\usepackage{graphicx,epstopdf,amsmath,amsfonts,amssymb,appendix,comment}
\usepackage{color,slashed,subfigure,setspace,footnote,multirow,longtable,braket}
\usepackage{epsfig,mathrsfs,latexsym,color,url,etoolbox}
\usepackage{bbm}
\definecolor{nicered}{rgb}{0.7,0.1,0.1}
\definecolor{nicegreen}{rgb}{0.1,0.5,0.1}
\definecolor{CarnationPink}{rgb}{1.0, 0.65, 0.79}
\DeclareMathAlphabet{\mathbbold}{U}{bbold}{m}{n}    

\usepackage{comment}

\usepackage{float}

\allowdisplaybreaks
\usepackage{dcolumn}% Align table columns on decimal point
\usepackage{bm}% bold math
\usepackage[table]{xcolor}% http://ctan.org/pkg/xcolor
\usepackage{multirow}
\usepackage{comment}
\usepackage{url}
\usepackage{tcolorbox}
\usepackage[T1]{fontenc}
\usepackage{cases}

\newcommand{\absq}[1]{\left\lvert #1 \right\rvert^2}
\def\Fermilab{Theoretical Physics Department, Fermilab, P.O. Box 500, Batavia, IL 60510, USA}
\def\Northwestern{Department of Physics \& Astronomy, Northwestern University, Evanston, IL 60208, USA}

\begin{document}

\preprint{FERMILAB-PUB-21-453-T, NUHEP-TH/21-11}

\title{Characterizing Heavy Neutral Fermions via their Decays}

\author[1]{Andr{\'e} de Gouv{\^e}a,}
\author[2]{Patrick J. Fox,}
\author[2]{Boris J. Kayser,}
\author[2]{Kevin J. Kelly}
\affiliation[1]{\Northwestern}
\affiliation[2]{\Fermilab}
\emailAdd{degouvea@northwestern.edu}
\emailAdd{pjfox@fnal.gov}
\emailAdd{boris@fnal.gov}
\emailAdd{kkelly12@fnal.gov}

\date{\today}% It is always \today, today,

\abstract{Many extensions of the Standard Model of particle physics contain new electrically-neutral fermions.  Should one of these particles be discovered, questions will naturally arise regarding its nature. For instance: is it a self-conjugate particle (i.e., is it a Dirac or a Majorana fermion)?, does it interact via the Standard Model force carriers or something else? One set of well-motivated particles in this class are Heavy Neutral Leptons (HNLs), Standard Model gauge-singlet fermions that mix with the neutrinos and may be produced in meson decays. We demonstrate that measuring the three body decays of the HNL (or phenomenologically similar heavy fermions) can help determine whether they are Majorana or Dirac fermions.  We also investigate the ability to distinguish among different models for the physics responsible for the HNL decay.  We compare the reach assuming full and partial event reconstruction, and propose experimental analyses.  Should a new fermion be discovered, studying its three body decays provides a powerful diagnostic tool of its nature.
}

\maketitle

%%%%%%%%%%%%%%%%%%%%%%%%%%%%%%%%%%%%%%%%%%%%
\section{Introduction}
\label{sec:Introduction}
%%%%%%%%%%%%%%%%%%%%%%%%%%%%%%%%%%%%%%%%%%%%

The existence of new, electromagnetically-neutral particles is often a prediction of new-physics scenarios aimed at addressing some of the outstanding contemporary questions in particle physics, including the dark-matter and neutrino-mass puzzles. A well-studied class of models contains fermions of this nature: the Heavy Neutral Leptons (HNLs)~\cite{Atre:2009rg,deGouvea:2015euy,Drewes:2015iva,Fernandez-Martinez:2016lgt,Drewes:2016jae,Bryman:2019ssi,Bryman:2019bjg,Bolton:2019pcu} (a wide range of experimental searches and proposals for HNLs are detailed in Refs.~\cite{Galeazzi:2001py,Hiddemann:1995ce,Belesev:2013cba,Holzschuh:1999vy,Holzschuh:2000nj,Derbin:1997ut,Schreckenbach:1983cg,Deutsch:1990ut,Britton:1992xv,Britton:1992pg,Aguilar-Arevalo:2017vlf,Bernardi:1985ny,Bernardi:1987ek,Abe:2019kgx,Baranov:1992vq,Bergsma:1985is,CooperSarkar:1985nh,Abreu:1996pa,Vaitaitis:1999wq,Artamonov:2009sz,Daum:1987bg,Hayano:1982wu,Coloma:2019htx,Arguelles:2019ziu,Abratenko:2019kez,Orloff:2002de,Coloma:2017ppo,Kobach:2014hea,Ballett:2016opr,Drewes:2018gkc,ArgoNeuT:2021clc,Kelly:2021xbv,Boiarska:2021yho,Arguelles:2021dqn}). While they are sometimes associated with other new particles and interactions\footnote{``HNL'' is typically used to denote a new, SM-gauge-singlet fermion whose only interactions with the SM field content is through a mass term that induces mixing with the light neutrinos. They have interactions with the SM $W$- and $Z$-bosons that are suppressed by a mixing angle. When considering a new fermion that has interactions beyond these, we will generically refer to them as ``heavy neutral fermions'' to reduce confusion. We will, however, use $N$ to represent the new particle in both cases.}, after gauge-symmetry breaking, HNLs mix with the standard model neutrinos. This mixing guarantees, independent from any other hypothetical interactions, that HNLs can be produced in charged-current and neutral-current processes. 

The discovery of HNLs or any other new neutral fermion would invite theoretical and experimental questions concerning their properties and their interactions. Among them, since they are fermions with zero electric charge, would be the nature of the new particles: Majorana fermions (MF) or Dirac fermions (DF)? The answer will, most likely, require a dedicated experimental effort. The type of the effort will depend on the mass of the fermions and their interactions. 

%\AdG{I find this paragraph a little out place. Would we benefit from demoting it to a footnote somewhere in the first paragraph?} Before proceeding, we clarify our language regarding such new-physics particles. The term ``Heavy Neutral Lepton'' is usually used in the literature to denote a new, SM-gauge-singlet fermion whose only interactions with the SM field content include a mass term (with the SM Higgs and lepton doublets) that induces mixing with the light neutrinos. Such particles then have interactions with the SM $W$- and $Z$-bosons that are suppressed by a mixing angle. When considering a new fermion that potentially has interactions beyond these (for instance, via some new light spin-0 and/or spin-1 bosons), we will generically refer to them as ``heavy neutral fermions'' to reduce confusion. However, we will generically use $N$ to represent the new particle in either case.

Since HNLs participate in charged-current interactions, if they are DF, they can be assigned the same lepton number as the standard-model neutrinos. Instead, if they are MF they will mediate lepton-number violating processes.  We will be interested in situations where the interactions of the HNLs preserve lepton number and any breaking of lepton number occurs softly, through mass terms.  Consider, for example, an HNL with a mass of order 100~GeV which can be produced in high-energy hadron collisions through the exchange of an off-shell $W$-boson, $(W^+)^*\to\ell^++N$, where $\ell$ are charged-leptons and $N$ is the HNL. If the HNL is a DF, it is constrained to decay, via the charged-current weak interactions, to a final state with lepton-number $+1$, e.g., $N\to W^+\ell^-$.  Instead, if the HNL is a MF, the lepton number of the daughters of the $N$-decay are not fixed, e.g., both $N\to W^+\ell^-$ and $N\to W^-\ell^+$ are allowed. This implies that MF HNLs will mediate processes that explicitly violate lepton-number, e.g., $pp\to (W^+)^*+$~hadrons~$\to \ell^+\ell^++$~hadrons. In this case, if the HNL decay is fast enough, it is possible to establish, on an event-by-event basis, that lepton number is not conserved and the HNL is a MF. 

\begin{figure}
\begin{center}
\includegraphics[width=0.9\linewidth]{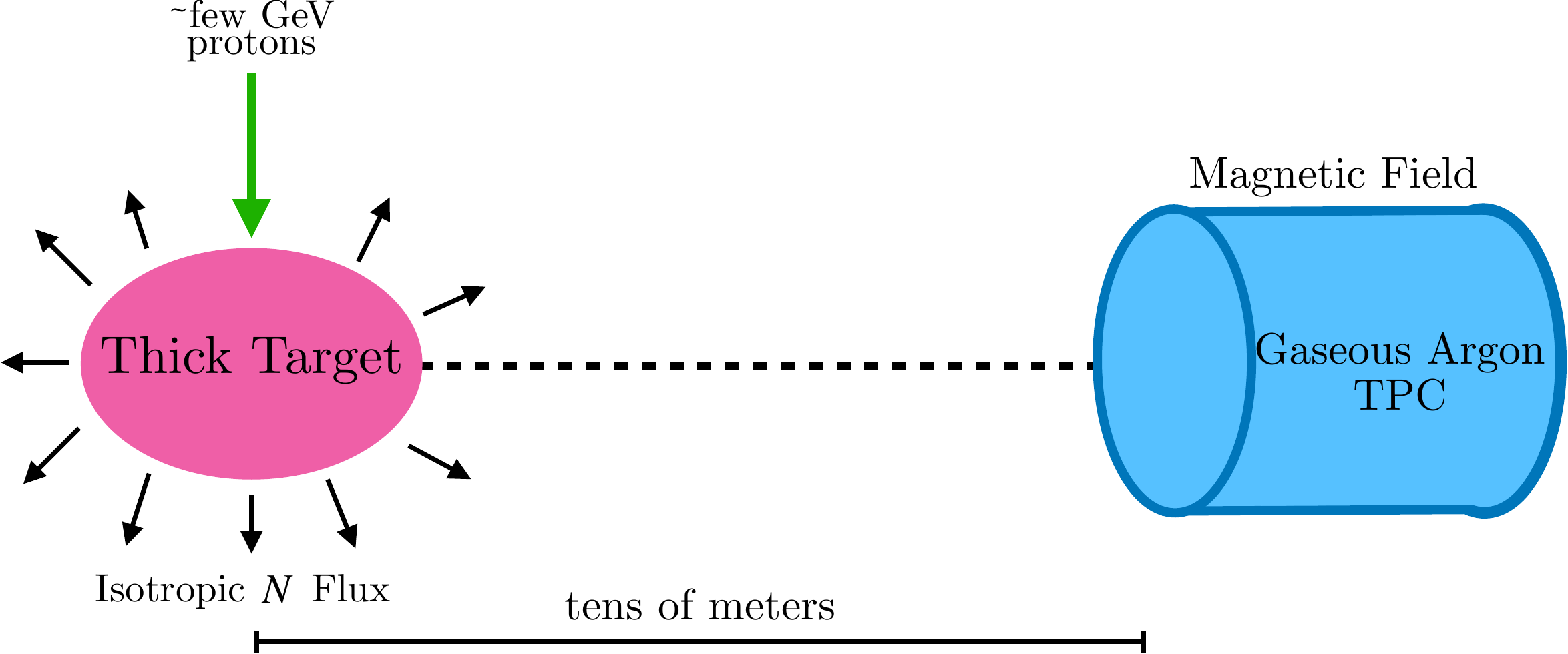}
\caption{Proposed setup for a post-discovery experiment to precisely study the nature of the new heavy fermion $N$. Protons with ${\sim}$few GeV energy strike a thick target, producing a large flux of charged mesons. Negatively-charged mesons are absorbed before they can decay and positively-charged ones are stopped and decay at rest. If the mesons decay into $N$, then we expect an isotropic flux of $N$. Some fraction will travel in the direction of the proposed gaseous argon time projection chamber and decay within, producing striking signals; the daughter particles can be measured precisely. 
\label{fig:ExpSetup}}
\end{center}
\end{figure}

For lighter new-physics fermions, the situation is qualitatively different. In what follows, we are interested in particles with masses between 1~MeV and 1~GeV. Given existing constraints on the new particles' existence, these are outside the reach of collider experiments but are the subject of fixed-target-like experiments. In more detail, the new fermions are produced in rare two- and three-body weak decays of mesons and later decay, after escaping the target-region, into light mesons, charged-leptons and neutrinos. Concretely, we assume the experimental setup depicted in Fig.~\ref{fig:ExpSetup}: an intense beam of ${\sim}$few~GeV protons strikes a thick target, producing a large flux of mesons. These are either captured or stopped in the target and decay at rest into charged-leptons and new fermions. The new fermions find their way to a gaseous argon time projection chamber and decay in such a way that the properties of their daughters, except for light neutrinos, are measured precisely.  

Here, it is either very difficult or outright impossible to establish, on an event-by-event basis, whether the new particle mediates lepton-number violating processes, for a couple of reasons. The first reason is that the lepton-number of the initial state is often unknown. In a thick target experiment, any particle (or particles) produced together with the new fermion in meson decay is absorbed by the target before its charge can be determined.\footnote{The exception to this is a ``tagged'' HNL beam, where the charge of the companion-charged-lepton is measured on an event-by-event basis.} This obstacle can be overcome statistically. In a thick target environment, negatively-charged mesons are less abundant than positively-charged mesons and are frequently absorbed by the target nuclei before they can decay. Positively-charged mesons are much more likely to decay after stopping. Hence, if the particle is a DF, one expects a ``beam'' that is predominantly made of $N$ rather than $\overline{N}$. The second reason is that, in the context of HNL phenomenology, light HNLs decay into final states involving light neutrinos a substantial (albeit model-dependent) fraction of the time. Since the HNLs are exclusively measured via their decay-products, in this case their lepton-number cannot be determined on an event-by-event basis. This obstacle may be overcome by exploring the decays of the HNL into charged-leptons and hadrons. In this case, however, the detector needs to be able to measure the charge of the daughter-charged-lepton. If the detector is not magnetized, for example, this may be impossible on an event-by-event basis.  

There are ways to distinguish MF from DF that do not rely on the observation of explicit lepton-number violation. These are the main focus of this manuscript. The production and decay kinematical properties of neutral fermions depend on whether they are MF or DF. For example, MF HNLs produced in the process $e^+ e^- \to Z\to N \nu$ are forward-backward symmetric in the center-of-mass frame independent of the polarization of the electron and positron beams while DF are produced, in general, with a non-zero forward-backward asymmetry~\cite{Blondel:2021mss}. The decay of a MF HNL into a self-conjugate boson and a MF neutrino is isotropic in the rest-frame of the HNL, independent of the polarization of the fermion~\cite{Balantekin:2018azf,Balantekin:2018ukw}, while that of a DF, in general, is not. The kinematics of three-body decays of MF and DF are also, in many cases, qualitatively different, as we explored in great detail in Ref.~\cite{deGouvea:2021ual}.

In Section~\ref{sec:Formalism}, we review the kinematical properties of two-body and three-body decays of MF and DF. We allow for the decays to be mediated by an arbitrary set of heavy new physics particles. %i.e.~ we consider all possible dimension-6 operators.  
In Section~\ref{sec:Experiment}, we introduce the details of the experimental setup we simulate, summarized above, and discuss the production of MF and DF, including their polarization. In Section~\ref{sec:DirVsMaj}, we discuss how well one can distinguish the MF and DF hypotheses as a function of the amount of data available. In Section~\ref{sec:InteractionStructure}, assuming the nature of the fermion is known, we investigate how well the properties of the physics responsible for its decay can be measured. We provide some parting thoughts in Section~\ref{sec:Conclusions} and detail the statistical methods we use in Appendix~\ref{app:Unbinned}.

%%%%%%%%%%%%%%%%%%%%%%%%%%%%%%%%%%%%%%%%%%%%
\section{Formalism}
\label{sec:Formalism}
%%%%%%%%%%%%%%%%%%%%%%%%%%%%%%%%%%%%%%%%%%%%
Throughout this work, we will be interested in the decay distributions of heavy, neutral fermions $N$. We will assume $N$ to be (at least partially) polarized, so this distribution depends on two (five) final-state kinematical quantities for two-body (three-body) decays. In the following two subsections, we review existing results and clarify the notation we will use throughout this work. Section~\ref{sec:TwoBodyFormalism} presents this for two-body decays -- we direct the reader to Refs.~\cite{Balantekin:2018azf,Balantekin:2018ukw} for further detail. Section~\ref{sec:ThreeBodyFormalism} repeats this for the more complicated three-body decay scenario -- more detail can be found in Ref.~\cite{deGouvea:2021ual}. Unless otherwise specified, we will always consider decay distributions in the rest frame of the decaying particle $N$.

\begin{figure}
\includegraphics[width=\linewidth]{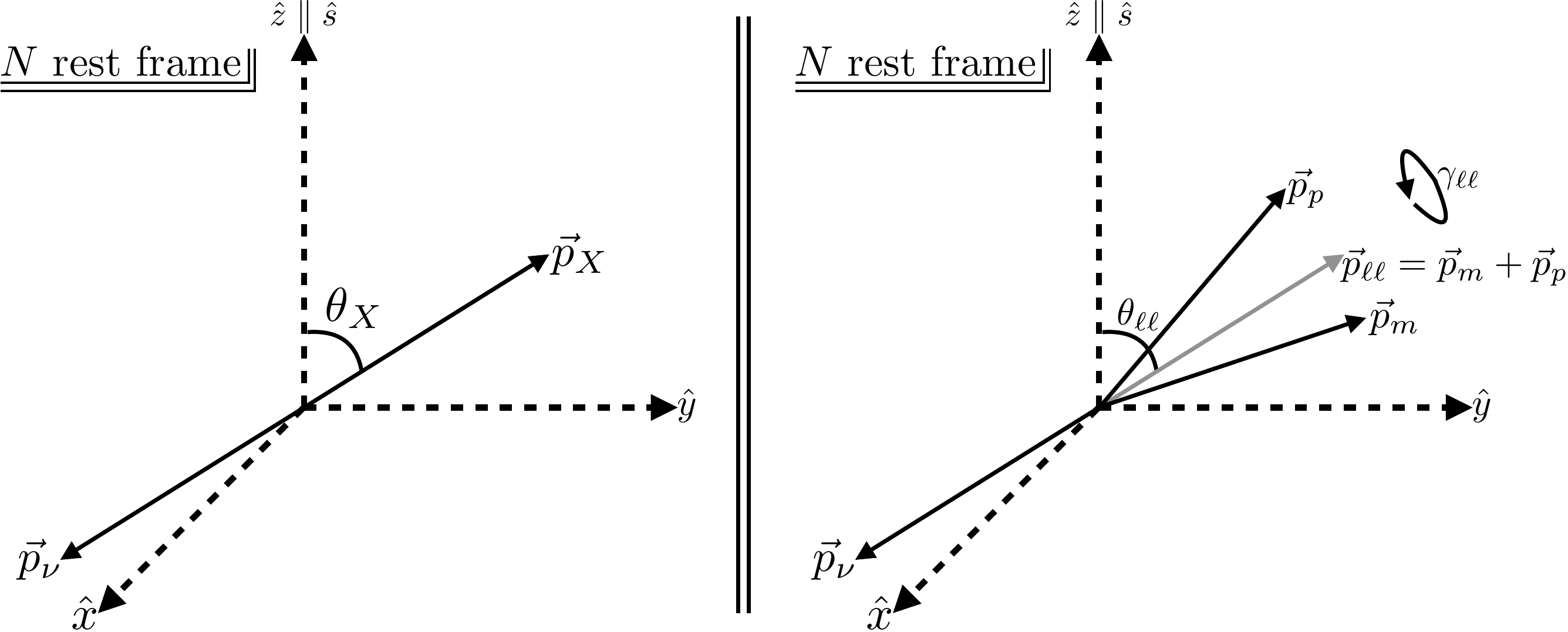}
\caption{Outgoing phase space of the decay of $N$ in its rest frame for two-body (left) and three-body (right) decays. Momenta of final state particles are labelled -- the two-body decay is the process $N \to \nu X$ and the three-body is $N \to \nu\ell_\alpha^- \ell_\beta^+$, where $\vec{p}_m$ ($\vec{p}_p$) is the three-momentum of the negatively- (positively-)charged lepton. \label{fig:Kinematics}}
\end{figure}

\subsection{Two-Body Decays}\label{sec:TwoBodyFormalism}
The two-body decay of a polarized, spin-$1/2$ particle $N \to \nu X$ requires two kinematical quantities in order to define the phase space. One of these, a rotation angle $\phi$ about the spin direction, is ignorable. We choose the other to be $\theta_X$, the angle between the spin direction and the outgoing $X$ particle. The phase space is flat with respect to $\cos\theta_X$ and we can express the two-body differential width as
\begin{equation}
\frac{d\Gamma(N \to \nu X)}{d\cos\theta_X d\phi} = \frac{1}{32\pi^2} \frac{\left\lvert \vec{p}_X\right\rvert}{m_N} \left\lvert \mathcal{M}\right\rvert^2,
\end{equation}
where $\left\lvert \vec{p}_X\right\rvert$ can be determined from energy-momentum conservation and $\left\lvert \mathcal{M}\right\rvert^2$ is the matrix-element-squared of this decay, which is independent of $\phi$. If $N$ is unpolarized, then there is additionally no $\cos\theta_X$ dependence and the familiar $1/(8\pi)$ factor of a two-body decay is recovered.

Refs.~\cite{Balantekin:2018azf,Balantekin:2018ukw} explored these processes in some detail, considering the cases where $X$ is a neutral Standard Model boson. As we will discuss later, objects including the forward-backward asymmetry of these two-body decays can be useful in differentiating between the possibilities that $N$ is either a DF or a MF.

\subsection{Three-Body Decays}\label{sec:ThreeBodyFormalism}
In contrast to the two-body case, the three-body decay of a polarized $N$ into the three-body final state $\nu \ell_\alpha^- \ell_\beta^+$ depends on five quantities. As above, the matrix-element-squared is independent of one of these (a rotation angle $\phi$ about the $N$ spin axis). We define $p_m^\mu$ ($p_p^\mu$) to be the four-momentum of the negatively- (positively-)charged lepton, and $p_{\ell\ell}^\mu \equiv p_m^\mu + p_p^\mu$. The remaining four kinematical quantities can be chosen to be
\begin{itemize} 
\item $m_{\ell\ell}^2 \equiv \left(p_{\ell\ell}^\mu\right)^2$, the invariant mass squared of the charged lepton pair. We will often use the dimensionless $z_{\ell\ell} \equiv m_{\ell\ell}^2/m_N^2$ in what follows;
\item $m_{\nu m}^2 \equiv \left(p_m^\mu + p_\nu^\mu\right)^2$, the invariant mass squared of the neutrino/negatively-charged lepton system.\footnote{This observable can only be measured if the charge of each final-state charged lepton can be determined -- $m_{\nu m}^2$ can be calculated by measuring the positively-charged lepton's energy in the $N$ rest frame.} Similarly, we will often use $z_{\nu m} \equiv m_{\nu m}^2/m_N^2$;
\item $\cos\theta_{\ell\ell}$, the angle between the $N$ spin-direction and $\vec{p}_{\ell\ell}$;
\item $\gamma_{\ell\ell}$, the angle of rotation of the charged-lepton subsystem about the direction of $\vec{p}_{\ell\ell}$.
\end{itemize}
The directions of these outgoing particles are shown in Fig.~\ref{fig:Kinematics}(right), modified from Ref.~\cite{deGouvea:2021ual}. While this set seems complicated, it is chosen so that the phase space is flat and the differential decay can be expressed as
\begin{equation}\label{eq:diffxsec}
\frac{d\Gamma\left(N \to \nu \ell_\alpha^- \ell_\beta^+\right)}{d \cos\theta_{\ell\ell} d\gamma_{\ell\ell} d m_{\ell\ell}^2 d m_{\nu m}^2 d\phi} = \frac{1}{\left(2\pi\right)^5} \frac{1}{64 m_N^3} \left\lvert \mathcal{M}\right\rvert^2,
\end{equation}
where, as above, $\left\lvert \mathcal{M}\right\rvert^2$ is the matrix-element-squared of this process.

Ref.~\cite{deGouvea:2021ual} presented calculations of this decay for a number of scenarios, specifying the form(s) that $\left\lvert \mathcal{M}\right\rvert^2$ takes for the most general possible dimension-6 interaction structure. Additionally, the forward-backward asymmetry (with respect to the direction $\theta_{\ell\ell}$) can be useful for discriminating between the Dirac/Majorana fermion hypotheses, as well as for determining the interaction structure of the $N$ decay. We will quantify how useful this information is in the coming sections.

We will rely on the notation defined in Ref.~\cite{deGouvea:2021ual} regarding the types of interactions that mediate the three-body decay $N \to \nu \ell_\alpha^- \ell_\beta^+$, allowing this four-fermion interaction to be as generic as possible. The matrix element $\mathcal{M}$, assuming any new mediators have been integrated out, can be written in terms of a ``neutral-current'' spinor ordering, 
\begin{equation}\label{eq:MGNL}
\mathcal{M} = G_{NL} \left[ \overline{u}_\nu \Gamma_N P_S u_N\right] \left[ \overline{v}_\beta \Gamma_L u_\alpha\right].
\end{equation}
Here, $\Gamma_N$ and $\Gamma_L$ represent different Lorentz structures of the scalar, pseudoscalar, vector, axial-vector, and tensor types, and $P_S$ is a spin-projection operator for the polarized $N$. The constants $G_{NL}$ represent different contributions to this process, and there are in total eighteen free parameters (each $G_{NL}$ can be complex):
\begin{equation}
\left\lbrace G_{SS}, G_{SP}, G_{PS}, G_{PP}, G_{VV}, G_{VA}, G_{AV}, G_{AA}, G_{TT} \right\rbrace.
\end{equation}
A set of couplings, which may be related to $G_{NL}$ if $\alpha = \beta$, can also contribute to decays of DF $\overline{N}$ or MF $N$ via the matrix element
\begin{equation}\label{eq:MbGNL}
\overline{\mathcal{M}} = \overline{G}_{NL} \left[ \overline{v}_N P_S \Gamma_N v_\nu\right] \left[ \overline{u}_\alpha \Gamma_L v_\beta\right].
\end{equation}
In what follows, we will be focusing on decays where $\alpha = \beta$. Ref.~\cite{deGouvea:2021ual} found that, for MF decays into identical final-state charged leptons, only the real or imaginary part of $G_{NL} \pm \overline{G}_{NL}$ enters the final matrix-element-squared (where the difference between real/imaginary and plus or minus depends on which pair of $N$, $L$ is considered).

Ref.~\cite{deGouvea:2021ual} explored the ramifications of all of these contributions to the decays of DF and MF $N$. For instance, when $N$ is a MF or when the final-state charged leptons are identical ($\alpha=\beta$), certain terms vanish. We will use this framework throughout the remainder of this work when discussing three-body decays $N \to \nu \ell_\alpha^- \ell_\beta^+$.

\subsection{Forward-Backward Asymmetry}\label{sec:FBAsymm}
One quantity of interest is the forward-backward asymmetry of a decay $A_{\rm FB}$. Such a quantity requires defining with respect \emph{to what} the forward-backward asymmetry is measured: we take this to be the direction of the outgoing $X$ (charged lepton pair) in the two-body (three-body) case relative to the spin of the decaying particle, as measured in the $N$ rest frame. More concretely,
\begin{equation}
A_{\rm FB} \equiv \displaystyle\frac{\Delta\Gamma}{\Gamma} = \displaystyle\frac{\displaystyle\int_0^1 \displaystyle\frac{d\Gamma}{d\cos\theta_X} d\cos\theta_X - \int_{-1}^{0} \displaystyle\frac{d\Gamma}{d\cos\theta_X} d\cos\theta_X}{\displaystyle\int_0^1 \displaystyle\frac{d\Gamma}{d\cos\theta_X} d\cos\theta_X + \int_{-1}^{0} \displaystyle\frac{d\Gamma}{d\cos\theta_X} d\cos\theta_X}.
\end{equation}
For the three-body case, $\theta_X$ is replaced by $\theta_{\ell\ell}$ and all other kinematical quantities are integrated over. Refs.~\cite{Balantekin:2018azf,Balantekin:2018ukw} proved that $A_{\rm FB} = 0$ for two-body decays of MF $N$ of this type, and Ref.~\cite{deGouvea:2021ual} explored how $A_{\rm FB}$ can be used to differentiate between three-body decays of DF and MF. In Section~\ref{subsec:AllowedAnisotropyPolarization} we discuss how these results are modified in light of a partially polarized $N$.

%%%%%%%%%%%%%%%%%%%%%%%%%%%%%%%%%%%%%%%%%%%%
\section{Post-discovery Experiment to Determine Heavy Fermion Nature}
\label{sec:Experiment}
%%%%%%%%%%%%%%%%%%%%%%%%%%%%%%%%%%%%%%%%%%%%
We are interested in the following possible future scenario: a new heavy fermion in the mass range $\mathrm{MeV} \lesssim m_N \lesssim \mathrm{GeV}$ is discovered in one or more experiments, and that the subsequent goal is to design a follow-up experiment to better understand the properties of this new particle -- does it have interactions in addition to mixing with the standard-model neutrinos? Is it a DF or a MF? To what final-states can it decay, and with what branching ratios?

We assume that, with detection in a current/near-future experiment, we can determine the mass of the newly-discovered $N$ with sufficient precision to hone in on its potential production mechanisms, and the kinematically-allowed final states to which it may decay. Since we are focused on three-body decays into a light neutrino and a pair of charged leptons, we will restrict ourselves to the scenario that $m_N > 2m_e$. We will assume that $N$ is produced via some mixing with the light neutrinos, and therefore can emerge from charged meson decays, specifically those of $\pi^\pm$ and $K^\pm$. We refrain from considering $D_{(s)}^\pm$ and $B$ mesons as they are above the threshold of the beam energies we consider, and their prompt decay does not lead to HNL production through meson decay at rest.

Decay processes with distinct, charged, final-state particles, e.g. $N \to \mu^- \pi^+$ and $N \to \nu \mu^+ e^-$, are particularly useful for distinguishing between the DF and MF hypotheses for $N$ by using combined searches for the charge-conjugated final states. By measuring the relative rate of $\mu^- \pi^+$ and $\mu^+ \pi^-$ events (assuming a detector with charge identification), this MF/DF distinction can be tested, as explored in Refs.~\cite{Formaggio:1998zn,Gorbunov:2007ak,Asaka:2012bb,Ballett:2016opr,Coloma:2017ppo,Arbelaez:2017zqq,Cvetic:2018elt,Bondarenko:2018ptm,Curtin:2018mvb,SHiP:2018xqw,Ariga:2018uku,Krasnov:2019kdc,Abe:2019kgx,Ballett:2019bgd,Drewes:2019byd,Chun:2019nwi,Arguelles:2019ziu,Abratenko:2019kez,Tastet:2019nqj,Berryman:2019dme,Gorbunov:2020rjx,Coloma:2020lgy,Batell:2020vqn,deVries:2020qns,Plestid:2020ssy,Breitbach:2021gvv} among others. If we consider final states with indistinguishable charged particles, e.g. $N \to \nu e^+ e^-$ and $N\to \nu \mu^+ \mu^-$, we must go beyond these rate measurements to probe the DF/MF distinction.

The remainder of this section is as follows. In Section~\ref{subsec:ProdPol}, we discuss how these $N$ would be produced in charged-meson decays, and how the resulting $N$ would be polarized when produced in this way. Ref.~\cite{deGouvea:2021ual} discussed how large the forward/backward asymmetries can be for MF and DF $N$ -- we combine this information and the polarization to see how large effects can be in a realistic experimental environment in Section~\ref{subsec:AllowedAnisotropyPolarization}. Section~\ref{subsec:DecaySignatures} discusses the different decay signatures of $N$ of interest for us, expanding on some of the arguments above and motivating our interest in $N \to \nu \ell^\pm \ell^\mp$. Finally, Section~\ref{subsec:Detector} introduces our proposed detector for such searches.

%%%%%%%%%%%%%%%%%%%%%%%%%%%%%%%%%%%%%%%%%%%%
\subsection{Production and Polarization of Heavy Fermions}
\label{subsec:ProdPol}
%%%%%%%%%%%%%%%%%%%%%%%%%%%%%%%%%%%%%%%%%%%%
$N$ will be produced in the decays of charged mesons $\mathfrak{m} = \pi^\pm$, $K^\pm$ via mixing with the light neutrinos in two-body\footnote{Three-body decays, such as $K^+ \to \pi^0 \ell^+ N$, can also contribute to $N$ production, as studied in Refs.~\cite{Gorbunov:2007ak,Coloma:2020lgy}. These three-body decays are usually subdominant to the two-body ones we will consider.} decays. We assume that this mixing generates a branching fraction $\mathrm{Br}\left( \mathfrak{m} \to \ell_\alpha N\right)$ that is consistent with current experimental data. When considering the decay signatures of $N$ in the following subsections, we will assume that the new interactions that we introduce to generate those decays dominate over this mixing-based interaction in determining the lifetime of $N$ and its branching ratios.

When produced in two-body decays $\mathfrak{m}^+ \to \ell^+ N$, the $N$ can emerge from the decay preferentially polarized because the interaction governing this decay is parity violating. The degree of polarization of the outgoing $N$ is of critical importance if $N$ is a DF and we want to analyze its decays to determine whether it is a DF or MF. When exploring the forward/backward asymmetry of DF $N$ decays, Refs.~\cite{Balantekin:2018azf,Balantekin:2018ukw,deGouvea:2021ual} did so assuming perfectly polarized $N$. If the $N$ is not polarized, any distinction between the DF and MF hypotheses using this asymmetry will be reduced -- we explain this in more detail in Section~\ref{subsec:AllowedAnisotropyPolarization}.

We can determine the polarization degree $P$ by calculating the decay rate of the meson into a spin-polarized $N$ for which the matrix element is
\begin{equation}\label{eq:MMeson}
\mathcal{M}_\mathfrak{m} = \sqrt{2} G_F f_{\mathfrak{m}} V_{qq^\prime} k^\alpha \left[ \overline{u}_{N} P_S \gamma_\alpha P_L v_\ell\right],
\end{equation}
where $G_F$ is the Fermi constant, $f_{\mathfrak{m}}$ is the decay constant of $\mathfrak{m}$, $V_{qq^\prime}$ is the relevant CKM matrix element for this decay, $k^\alpha$ is the four-momentum of the parent $\mathfrak{m}$, $P_L$ is the left-chiral projection operator, and $P_S = \frac{1}{2}(1 + \gamma^5 \slashed{s})$ is the spin-projection operator. We define the spin to be in the direction of the outgoing $N$, $\vec{p}_N$ (all calculated in the rest-frame of $\mathfrak{m}$) such that
\begin{equation}
s^\mu = \frac{\lambda_N}{m_N} \left(\left\lvert \vec{p}_N\right\rvert; \hat{p}_N E_N\right).
\end{equation}
where $\lambda_N = 1\ (-1)$ corresponds to a right-handed (left-handed) outgoing $N$.

\begin{figure}
\begin{center}
\includegraphics[width=0.7\linewidth]{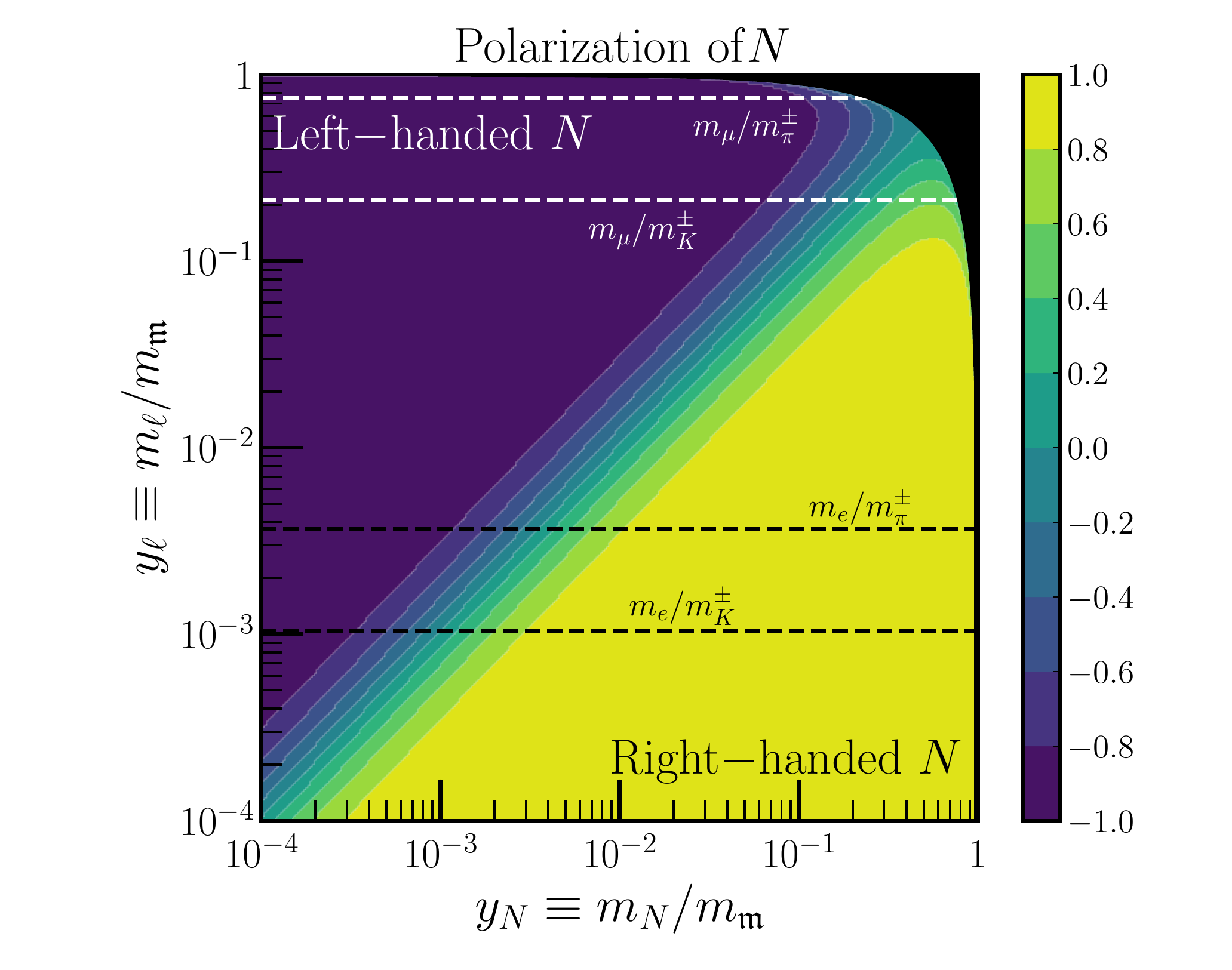}
\caption{Polarization of the outgoing $N$ from the decay $\mathfrak{m}^+ \to \ell^+ N$, with $y_N \equiv m_N/m_\mathfrak{m}$ and $y_\ell = m_\ell/m_N$. Yellow (purple) regions correspond to the outgoing $N$ preferring to be more right-(left)-handed in these decays. The black region is forbidden -- $y_N + y_\ell > 1$. Horizontal lines correspond to values of $y_\ell$ for muons and electrons emerging from charged pion and kaon decays, as labelled.
\label{fig:Polarization}}
\end{center}
\end{figure}

Next, we square the matrix-element in Eq.~\eqref{eq:MMeson} and calculate the decay widths when $\lambda_N = \pm 1$. The total width is then obtained when we sum the two possibilities. This gives us the polarization degree $P$,
\begin{equation}
P \equiv \frac{\Gamma_{\lambda_N = +1} - \Gamma_{\lambda_N = -1}}{\Gamma_{\lambda_N = +1} + \Gamma_{\lambda_N = -1}},
\end{equation}
where $P \to 1$ if the $N$ always come out right-handed (in the $\mathfrak{m}$ reference frame) and $P \to -1$ if they always come out left-handed. Defining the ratios of masses $y_\ell \equiv m_\ell/m_\mathfrak{m}$ and $y_N \equiv m_N/m_\mathfrak{m}$, we find that
\begin{equation}\label{eq:Pol}
P(y_\ell, y_N) = \frac{\left(y_N^2 - y_\ell^2\right) \sqrt{ \left(1 - y_\ell^2\right)^2 - 2y_N^2 \left(1 + y_\ell^2\right) + y_N^4}}{\left(1- y_\ell^2\right) y_\ell^2 + 2y_\ell^2 y_N^2 + \left(1 - y_N^2\right)y_N^2}.
\end{equation}
Fig.~\ref{fig:Polarization} displays the value of $P$ as a function of $y_\ell$ and $y_N$.
Yellow (purple) are the regions where $P \to 1$ ($-1$), meaning that the $N$ emerge preferentially right-(left-)handed. The black region indicates where $y_\ell + y_N > 1$ which is kinematically forbidden. Dashed, horizontal lines in Fig.~\ref{fig:Polarization} indicate values of $y_\ell$ for charged meson decays into muons and electrons (along with $N$) for charged pion and kaon decays. The polarization $P$ goes to zero when the $N$ and the outgoing charged-lepton in the two-body decay $\mathfrak{m} \to \ell N$ are of the same mass, see Eq.~\eqref{eq:Pol}, which means it is not possible to produce polarized $N$ when they are mass-degenerate with the corresponding SM charged lepton.

%Going forward, we will consider the following experimental setup for our hypothetical post-discovery experiment. We assume that a large quantity of charged mesons $\mathfrak{m}^\pm$ are produced when an intense beam of energetic protons strikes a target. We assume that this target is high-$Z$ \KJK{Check this logic}, allowing for the negatively-charged mesons $\mathfrak{m}^-$ to be absorbed by the nuclei before decaying, and that the $\mathfrak{m}^+$ are stopped very quickly and decay at rest in this two-body way.

%For concreteness, we assume that the production is driven by a JSNS-like source, a $\mathrm{O}(3\ \mathrm{GeV})$ proton beam striking a mercury target. JSNS is capable (at 1 MW power) of delivering $3.8 \times 10^{22}$ protons on target per year, and per proton-on-target, roughly $0.01$ $K^+$ and $0.3$ $\pi^+$ are produced. If we want to consider $D_{(s)}^+$ production, we need considerably higher beam energy. The decision of what beam energy to use will depend strongly on the initial discovery of $N$ -- we will assume henceforth that it is identified that $N$ is likely produced in $\pi^+$ and/or $K^+$ decays. A schematic representation of this type of production, and the resulting isotropic $N$ flux, is shown in Fig.~\ref{fig:ExpSetup}.

\subsection{Allowed Anisotropy with Polarization}\label{subsec:AllowedAnisotropyPolarization}
Ref.~\cite{deGouvea:2021ual} explored how large the forward/backward asymmetry of three-body MF and DF decays $N \to \nu \ell_\alpha^- \ell_\beta^+$ can be under certain model assumptions for perfectly polarized $N$, parameterized by the asymmetry parameter $A_{\rm FB}$. If the source of $N$ is not perfectly polarized, such as in the situation discussed above, we must take into account this imperfect polarization when comparing to any sort of expected experimental signature.

As established in Ref.~\cite{deGouvea:2021ual}, we consider the matrix-element-squared for the three-body decay process (this result holds analogously for the two-body decays as well) as a dot product of spin-independent ($K_\alpha^\mathrm{I}$) or spin-dependent ($K_\alpha^\mathrm{D}$) Lorentz Invariants and a set of coefficients ($C^\alpha$), depending on the different couplings $G_{NL}$. 
We choose the $N$ to be right-handed i.e. $\hat{s} = \hat{p}_N$ and we denote the matrix-element-squared as $\left\lvert \mathcal{M}_R\right\rvert^2$. In order to consider left-handed decays (with the matrix-element-squared $\left\lvert \mathcal{M}_L\right\rvert^2$), all that must be done is to replace $\hat{s} \to -\hat{s}$ everywhere, or equivalently, flipping the sign of every spin-dependent Lorentz Invariant $K_\alpha^\mathrm{D} \to -K_\alpha^\mathrm{D}$.\footnote{This relies on the fact that all spin-dependent Lorentz Invariants that appear in our calculations are only linearly dependent on $s^\mu$, never quadratic or higher.} Above we defined the polarization degree $P$ and use it to determine the fraction of $N$ that are right- or left-handed: $f_{R,L} = (1\pm P)/2$ (such that $f_R + f_L = 1$). Then, we may include the effects of partial polarization by
\begin{align}
\frac{d\Gamma}{d\vec{\vartheta}} \propto \absq{\mathcal{M}} &= f_R \absq{\mathcal{M}_R} + f_L \absq{\mathcal{M}_L}, \nonumber\\
&= f_R \left(C^\alpha K_{R,\alpha}\right) + f_L \left(C^\alpha K_{L,\alpha}\right), \nonumber\\
&= f_R \left(C^\alpha \left(K_{\alpha}^{\mathrm{I}} + K_{\alpha}^{\mathrm{D}}\right)\right) +  f_L \left(C^\alpha \left(K_{\alpha}^{\mathrm{I}} - K_{\alpha}^{\mathrm{D}}\right)\right), \nonumber\\
&= C^\alpha \left( \left(f_R + f_L\right) K_\alpha^{\mathrm{I}} + \left(f_R - f_L\right) K_\alpha^{\mathrm{D}}\right), \nonumber\\
&= C^\alpha \left( K_\alpha^{\mathrm{I}} + PK_\alpha^{\mathrm{D}}\right). \label{eq:dGammaPol}
\end{align}
Effectively, this means that the imperfect polarization case may be understood by reweighting all spin-dependent terms by $P \in [-1,1]$.  

Now, we revisit some results of Ref.~\cite{deGouvea:2021ual} by including the factor $P$ in the figure of merit, and analyze how large $A_{\rm FB} \times P$ can be for some specific scenarios. For simplicity, we will only focus on the final-state decay $N \to \nu e^+ e^-$ and consider different production mechanisms for $N$. Since the identical final-state charged leptons imply that if $N$ is a MF, $A_{\rm FB} = 0$, we will only consider the effects here for DF $N$ decays.

In order for the decay $N \to \nu e^+ e^-$ to proceed, $m_N > 2 m_e$. This implies that, if we are considering production of the type $\mathfrak{m} \to e N$, the factors that enter Eq.~\eqref{eq:Pol} satisfy $y_N > 2y_\ell$. In this case, $N$ will be mostly right-handed, $P \approx 1$, for all masses of interest. This will have little impact on the overall forward-backward asymmetry so we will therefore focus on the combination of production and decay of DF $N$ where it is produced via $\mathfrak{m} \to \mu N$, and decays via $N \to \nu e^+ e^-$.

\begin{figure}
\begin{center}
\includegraphics[width=0.48\linewidth]{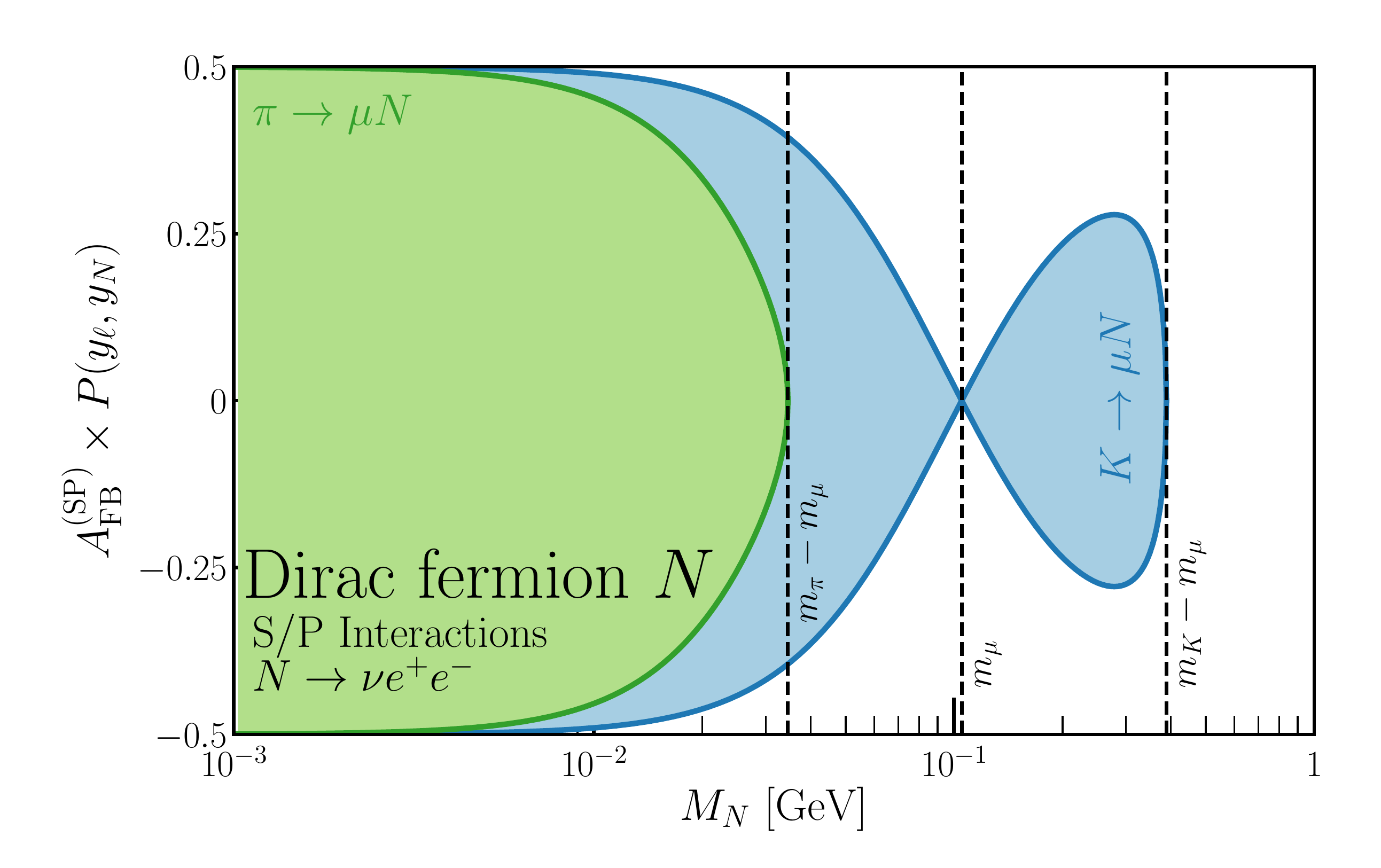}
\includegraphics[width=0.48\linewidth]{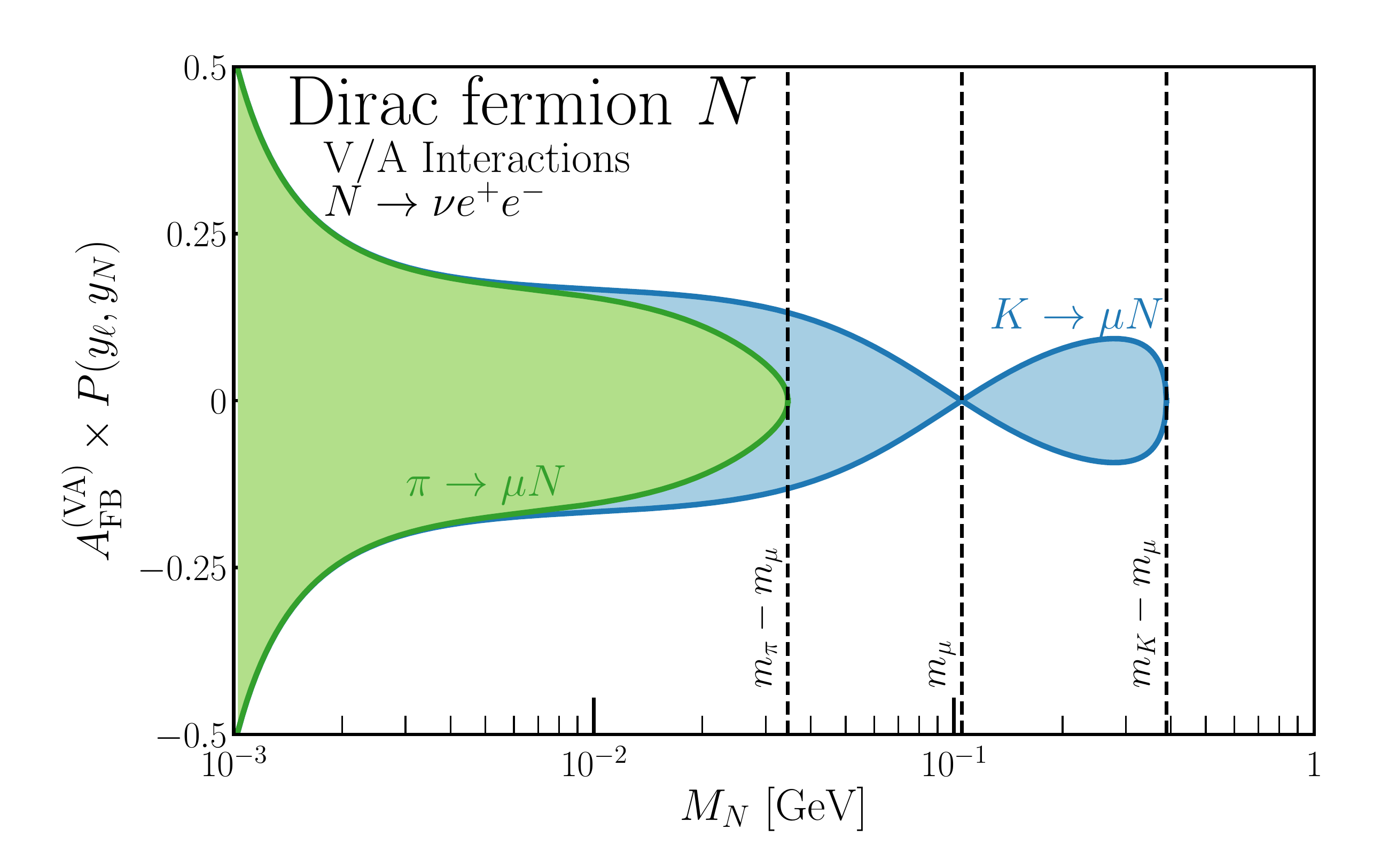}
\caption{Allowed effective asymmetry of the decay $N \to \nu e^+ e^-$ where $N$ is a Dirac fermion decaying with only Scalar/Pseudoscalar interactions (left) or only Vector/Axial-vector interactions (right). In each panel, we include polarization effects (see Eq.~\eqref{eq:Pol}) from the production of $N$, assuming it is generated in the two-body decay $\pi \to \mu N$ (green) or $K \to \mu N$ (blue). Dashed vertical lines indicate the kinematical endpoints of the two production mechanisms, as well as where $m_N = m_\mu$, where the polarization goes to zero.
\label{fig:AlphaPolDir}}
\end{center}
\end{figure}
If only Scalar/Pseudoscalar couplings are considered ($G_{SS}$, $G_{SP}$, $G_{PS}$, and $G_{PP}$ being the only nonzero couplings for $N$ decays), then the forward/backward asymmetry $A_{\rm FB}^{\rm (SP)}$ can be as large in magnitude as $1/2$, the largest allowed since the differential partial width is proportional to $(1+2A_{\rm FB}\cos\theta)$. Once partial-polarization effects are included, the largest effective asymmetry $A_{\rm FB}P$ is simply $P/2$. Fig.~\ref{fig:AlphaPolDir}(left) displays this for the two production channels of interest, $\pi \to \mu N$ and $K\to \mu N$. From Eq.~\eqref{eq:Pol} we see that $P \to 0$ when either $m_N \to m_\mathfrak{m} - m_\mu$ or when $m_N \to m_\mu$, explaining why the green and blue regions vanish at their respective kinematical endpoints and why the blue region pinches to a point when $m_N = m_\mu$.

The allowed forward-backward asymmetry for a perfectly-polarized DF $N$ decaying to $\nu e^+ e^-$ mediated by only vector and axial-vector interactions is more complicated, and the expressions for $A_{\rm FB}^{\rm (VA)}$ were given in Ref.~\cite{deGouvea:2021ual}, allowing for at most $A_{\rm FB}^{\rm (VA)} = 1/2$ for low $m_N$ and $A_{\rm FB}^{\rm (VA)} = 1/6$ for $m_N \gg 2m_e$. We now include the polarization $P$ to obtain the effective asymmetry -- see Fig.~\ref{fig:AlphaPolDir} (right). Again, we do so assuming $N$ is produced in the two-body decay $K \to \mu N$ (blue) or $\pi \to \mu N$ (green). The same features as above are present here: $P \to 0$ when $m_N \to m_\mathfrak{m} - m_\mu$ and when $m_N \to m_\mu$.

%%%%%%%%%%%%%%%%%%%%%%%%%%%%%%%%%%%%%%%%%%%%
\subsection{Signatures of Heavy Fermion Decay}
\label{subsec:DecaySignatures}
%%%%%%%%%%%%%%%%%%%%%%%%%%%%%%%%%%%%%%%%%%%%
If $N$ is in the mass range $\mathrm{MeV} \lesssim m_N \lesssim \mathrm{GeV}$, a large number of potential decay modes into SM particles are possible.  We divide these different decay modes into classes based on how useful they are in discriminating between the DF and MF hypotheses, and how one would go about performing such a search.

First, the ever-present but hopeless decay, $N \to 3\nu$. Unless one had an extraordinarily large flux of decaying $N$ and could see subsequent $\nu$ interactions, this channel is not of any use.

Next, we consider two-body decays of the type $N \to \nu X$, where $X$ is a self-conjugate boson. The structure of decays of this type was studied in detail in Refs.~\cite{Balantekin:2018azf,Balantekin:2018ukw} and reiterated in Section~\ref{sec:TwoBodyFormalism} -- we will quantify how useful they are in distinguishing between the DF and MF hypotheses in Section~\ref{subsec:TwoBodyDirMaj}. Ref.~\cite{Berryman:2019dme} previously determined that, in a neutrino-beam environment, these channels are likely background-dominated and have difficulties in reconstructing the $N$ rest frame. However, they could be of use in a decay-at-rest setting.

The third class are two-body decays of the type $N \to \ell^\pm \mathfrak{m}^\mp$. If $N$ is a DF then it will only decay into one of these final states, i.e., $N \to \ell^- \mathfrak{m}^+$, but if $N$ is a MF then it will decay into the two final states with equal probability. These channels are very useful for distinguishing between the MF and DF hypotheses (see, e.g., Ref.~\cite{Berryman:2019dme}). While the angular distributions of the outgoing $\ell^\pm$ and $\mathfrak{m}^\mp$ can be analyzed to separate between the hypotheses, the best way to distinguish the hypotheses is to measure the relative rates of $\ell^+ \mathfrak{m}^-$ and $\ell^- \mathfrak{m}^+$. This class includes the channels $N \to e^\pm \pi^\mp$, $N \to \mu^\pm \pi^\mp$, $N \to e^\pm K^\mp$, $N \to \mu^\pm K^\mp$.

Finally, we come to the three-body leptonic decays $N \to \nu \ell_\alpha^+ \ell_\beta^-$. As discussed in Ref.~\cite{deGouvea:2021ual}, when $\alpha \neq \beta$, measuring the decay distribution could help in separating the DF hypothesis from the MF one, but simply counting the relative rates of the two combinations $\ell_\alpha^+ \ell_\beta^-$ and $\ell_\alpha^-\ell_\beta^+$ also provides a large amount of information. When $\alpha=\beta$, counting alone is no longer of use and one requires the decay distribution in order to address this question. This class contains the channels $N \to \nu e^+ e^-$, $N \to \nu \mu^\pm e^\mp$, and $N \to \nu \mu^+ \mu^-$ (as well as final states involving $\tau^\pm$ if $N$ is heavy enough  --- given our boundary conditions $m_N \lesssim 1$ GeV, taus are never allowed).

If a particle $N$ is discovered in the ${\sim}1-100\ \mathrm{MeV}$ mass range, its only kinematically-accessible decays are $N \to 3\nu$, $N \to \nu\gamma$, and $N \to \nu e^+ e^-$. If the $N$ decay is governed by the weak interactions, $N \to \nu e^+ e^-$ can still have a sizeable branching ratio for significantly larger values of $m_N$. Generically, $\mathrm{Br}(N \to \nu e^+ e^-)$ is no smaller than $\mathcal{O}(10^{-2})$ for three orders of magnitude in $m_N$~\cite{Gorbunov:2007ak}. Therefore, we find the possibility of discovering an HNL decaying in this channel, and determining whether it is a Dirac or Majorana fermion, particularly intriguing.

%%%%%%%%%%%%%%%%%%%%%%%%%%%%%%%%%%%%%%%%%%%%
\subsection{Hypothetical Detector}
\label{subsec:Detector}
%%%%%%%%%%%%%%%%%%%%%%%%%%%%%%%%%%%%%%%%%%%%
In this sub-section we will consider a hypothetical future detector well-suited to search for the charged-lepton final states. Due to its low-momentum particle thresholds, excellent identification of electron and muon tracks, and charge-identification, we suggest the use of a gaseous argon time projection chamber, similar to the one planned for part of the Deep Underground Neutrino Experiment (DUNE) Near Detector~\cite{DUNE:2021tad}, for such a search. While the detector planned for DUNE must meet neutrino-oscillation-related requirements, which dictate its mass and density, we are not confined to such restrictions here. Because we want to search for decays, it is advantageous if the pressure is as low as possible (to suppress scattering-related backgrounds), which enables even lower-energy particle thresholds. We assume that this hypothetical detector, like the one planned for DUNE, is embedded inside an electromagnetic calorimeter that also has a muon-tagger (to reduce any confusion between muons and pions), all of which is situated in a magnetic field that allows for nearly 100\% charge identification.
Furthermore, we will carry out our analyses under the simplifying assumption of zero backgrounds.  Backgrounds can be greatly mitigated by placing the detector a distance of ${\sim}$tens of meters from the target so that backgrounds such as neutrons can be suppressed by surrounding dirt, rock, and (potentially) added shielding. Additionally, if  the detector is located perpendicular to the incident beam direction, backgrounds are further reduced, since the signal comes from isotropic decay at rest while many of the potential backgrounds are in line with the beam.

\begin{figure}[!t]
\begin{center}
\includegraphics[width=0.5\linewidth]{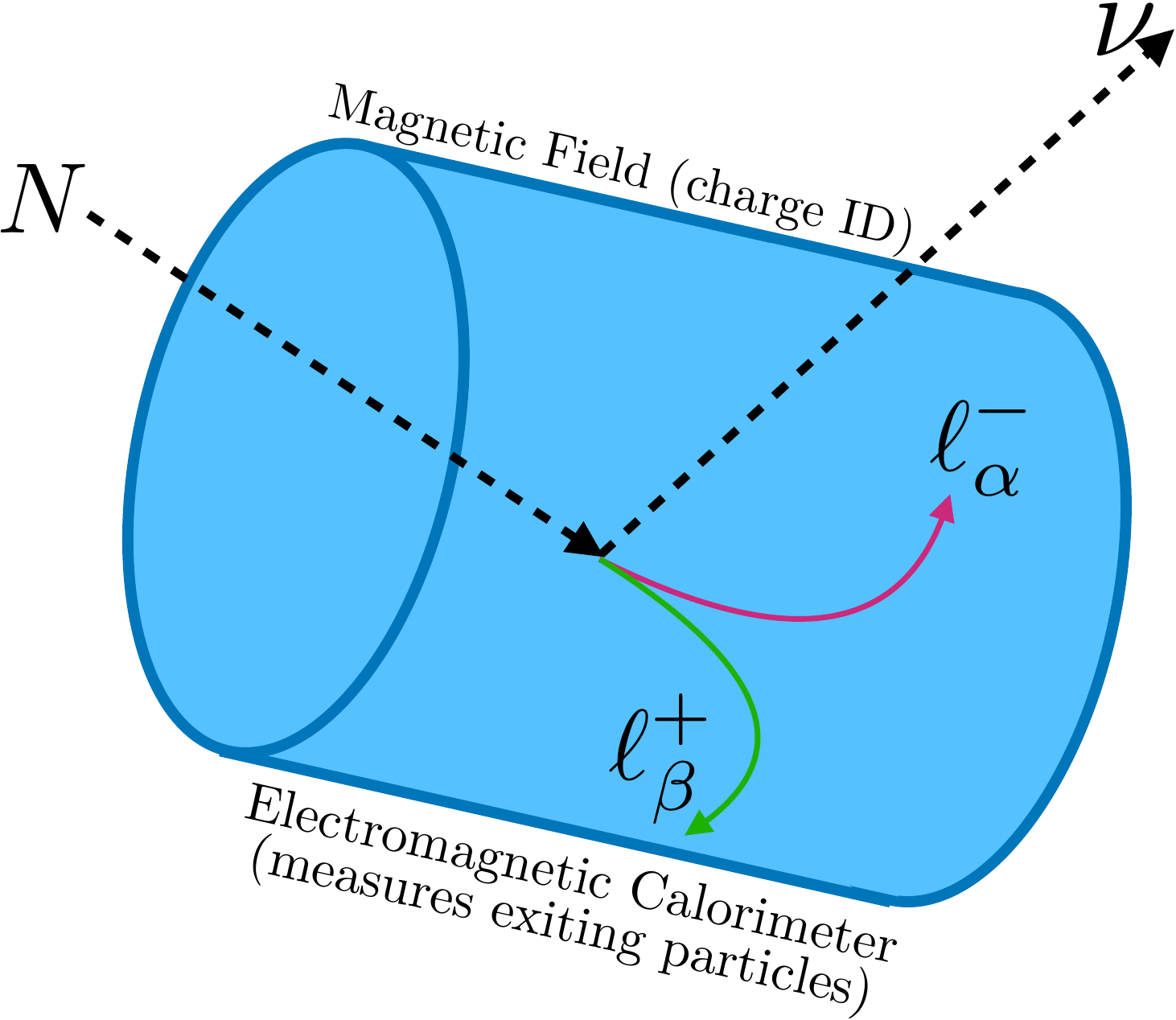}
\caption{Schematic representation of the decay $N\to \nu \ell_\alpha^- \ell_\beta^+$ occurring within the proposed gaseous argon time projection chamber. The magnetic field causes the charged lepton tracks to bend and have their charges identified, whereas the outgoing light neutrino escapes undetected. If the charged leptons exit the gaseous argon before coming to rest, their energy is measured in the electromagnetic calorimeter surrounding the time projection chamber.
\label{fig:DecaySketch}}
\end{center}
\end{figure}

Fig.~\ref{fig:ExpSetup} displays a sketch of the proposed setup described here and in Section~\ref{subsec:ProdPol}. In Fig.~\ref{fig:DecaySketch}, we show a mock version of the type of signal in this detector. 
Here, the decay $N \to \nu \ell_\alpha^- \ell_\beta^+$ is depicted within the gaseous argon time projection chamber, where the red (green) track indicates the $\ell_\alpha^-$ ($\ell_\beta^+$) depositing energy as it travels through the detector. The magnetic field bends the charged tracks, allowing us to identify which lepton has which charge nearly 100\% of the time. Any lepton that exits the detector before coming to rest will deposit its remaining energy in the surrounding electromagnetic calorimeter.

For the purposes of the remaining study, we will assume that this detector can measure the energies and directions of the outgoing charged leptons perfectly~\cite{DocDB}, as well as identify the vertex from which they emerged. If the mass $m_N$ is known, and $N$ is produced in a two-body decay-at-rest process, then its energy is uniquely determined. This allows us to transform from the laboratory quantities of the charged lepton energies/directions into the $N$ rest frame. The calculations of Ref.~\cite{deGouvea:2021ual}, restated in Section~\ref{sec:ThreeBodyFormalism}, use the rest frame kinematics. 

%%%%%%%%%%%%%%%%%%%%%%%%%%%%%%%%%%%%%%%%%%%%
\section{Distinguishing Between the Dirac Fermion and Majorana Fermion Hypotheses}
\label{sec:DirVsMaj}
%%%%%%%%%%%%%%%%%%%%%%%%%%%%%%%%%%%%%%%%%%%%
Above, we motivated searches for signatures of decays of $N$ that do \textit{not} have identifiable lepton number, i.e. those with one or more neutrinos in the final state. In this section, we explore how these searches can be used to distinguish between the DF and MF hypotheses for $N$. We focus on two-body decays of the type $N \to \nu X^0$ in Section~\ref{subsec:TwoBodyDirMaj}, extending on the work of Refs.~\cite{Balantekin:2018azf,Balantekin:2018ukw} to quantify the required number of signal events to perform this distinction. In Section~\ref{subsec:ThreeBodyDirMaj} we extend this approach to the three-body decays $N \to \nu \ell^+ \ell^-$. We do this under two assumptions. First, in Section~\ref{subsec:MFDF:YesSpin} we assume that $N$ is (at least partially) polarized and that we can use all relevant kinematical information to compare the DF and MF hypotheses. Then, in Section~\ref{subsec:MFDF:NoSpin}, we study the case where spin-related observables cannot be used. We find that, in some cases, some information for comparing the DF and MF hypotheses is still present in the absence of polarized $N$.

%%%%%%%%%%%%%%%%%%%%%%%%%%%%%%%%%%%%%%%%%%%%
\subsection{Two-Body Decays}\label{subsec:TwoBodyDirMaj}
%%%%%%%%%%%%%%%%%%%%%%%%%%%%%%%%%%%%%%%%%%%%
In this subsection, we explore the two-body decays $N\to\nu X^0$ described in Section~\ref{sec:TwoBodyFormalism}, where $X^0$ is a standard model particle. In order to obtain the most optimistic results possible, we imagine a detector that can identify the outgoing $X^0$ and measure its momentum with perfect precision, and that we can reconstruct the rest frame of $N$ perfectly so that we can determine the (rest-frame) outgoing direction of $X^0$, $\cos\theta_{X}$. We also assume that there are no background events associated with this search. A realistic search of this type is expected to be significantly more challenging.

An experimental setup in which this procedure is more difficult is, for instance, that of a neutrino beam environment where the $N$ are highly boosted. If there is a distribution of $N$ boosts, then transforming between the laboratory and rest frames is nontrivial. Additionally, differences in distributions of $\cos\theta_X$ in the $N$ rest frame transform into differences in distributions of the lab-frame $X$ energy, and if the parent $N$ are highly boosted, these differences are harder to distinguish (see Refs.~\cite{Balantekin:2018ukw,Berryman:2019dme} for more details). In contrast, if the $N$ are produced in a two-body decay-at-rest process then they are mono-energetic and this frame transformation is simpler. The $N$ will also not be as highly boosted in such a scenario, making it more likely that such energy measurements are possible.

The decays of a DF can be described using two parameters -- $N_{\rm Evt.}$, the total number of expected signal events of this type and $PA_{\rm FB}$, the product of the $N$ polarization and the anisotropy parameter of this decay. Reproduced from Refs.~\cite{Balantekin:2018azf,Balantekin:2018ukw}, Table~\ref{tab:AnisotropyTwoBody} gives the expected anisotropy parameter for $X^0 = \gamma,\ \pi^0,\ \rho^0,\ Z^0,$ and $H^0$, assuming the decays are mediated by left-handed interactions, with the exception of $N\to\nu\gamma$, which is assumed to proceed via a transition magnetic dipole moment $\mu_N$ and an electric dipole moment $d_N$.
\begin{table}
\begin{center}
\caption{Anisotropy parameters of two-body decays $N \to \nu X^0$ for a variety of final-state standard model particles $X^0$. Reproduced from Refs.~\cite{Balantekin:2018azf,Balantekin:2018ukw}.
\label{tab:AnisotropyTwoBody}}
\begin{tabular}{|c||c|c|c|c|c|}\hline
Particle & $\gamma$ & $\pi^0$ & $\rho^0$ & $Z^0$ & $H^0$ \\ \hline \hline
$A_{\rm FB}$ & $\displaystyle\frac{\mathrm{Im}\left(\mu_N d_N^*\right)}{\left\lvert \mu_N\right\rvert^2 + \left\lvert d_N\right\rvert^2}$ & $\displaystyle\frac{1}{2}$ & $\displaystyle\frac{m_N^2 - 2m_{\rho^0}^2}{2\left(m_N^2 + 2m_{\rho^0}^2\right)}$ & $\displaystyle\frac{m_N^2 - 2m_Z^2}{2\left(m_N^2 + 2m_Z^2\right)}$ & $\displaystyle\frac{1}{2}$ \\ \hline
\end{tabular}
\end{center}
\end{table}

With these ingredients, we can simulate the expected distribution of events in $\cos\theta_X$ under different model assumptions.
\begin{figure}
\begin{center}
\includegraphics[width=0.6\linewidth]{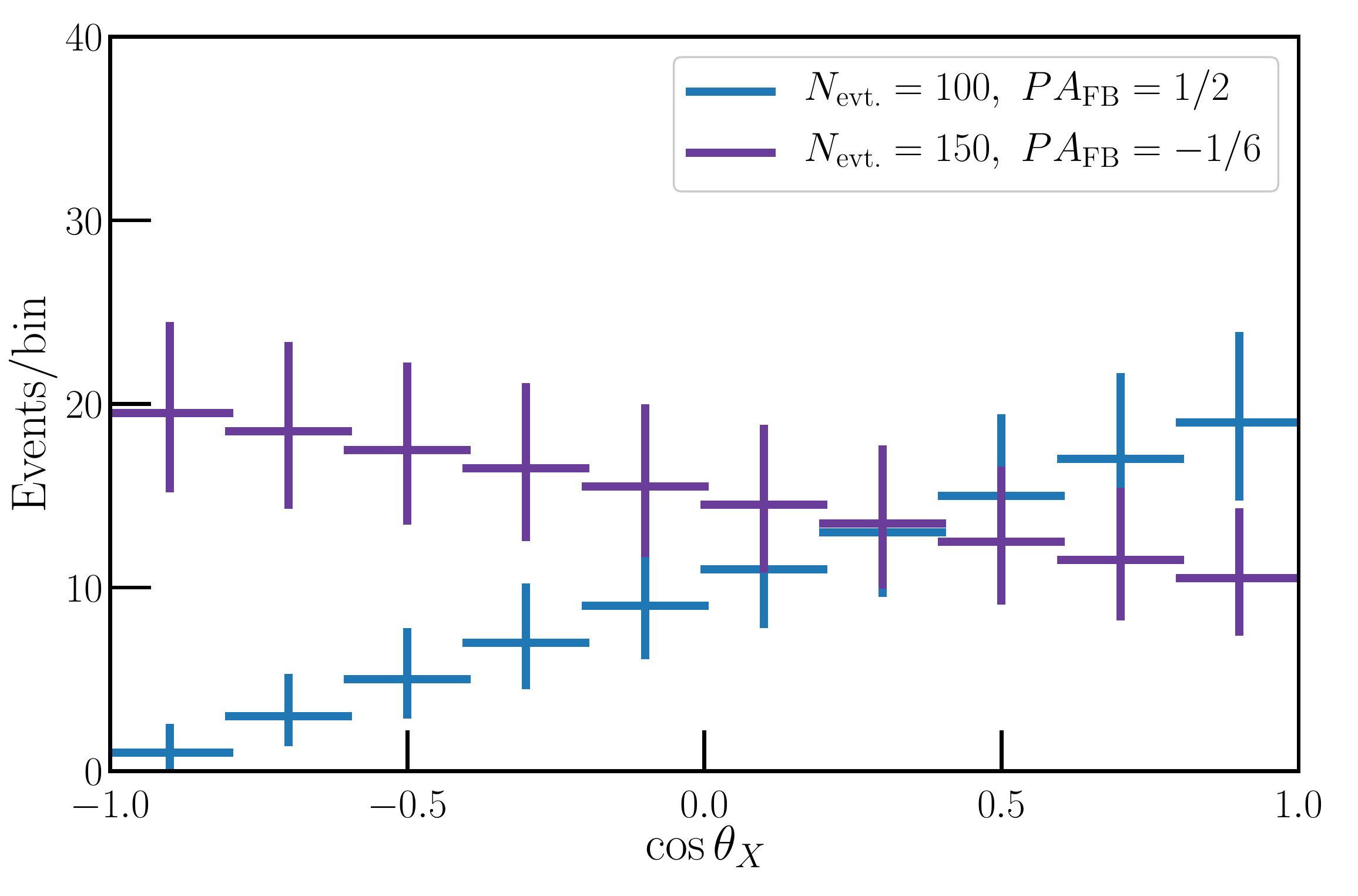}
\caption{Expected event distributions for two-body decays $N \to \nu X^0$ assuming $N$ is a Dirac Fermion with polarization/decay anisotropy as indicated in the legend. Purple points indicate a spectrum with a total expected number of events of $150$ and a small anisotropy/polarization, where the blue indicate a spectrum with $100$ expected events and maximal anisotropy/polarization. Error bars here are only statistical.
\label{fig:TwoBodyDist}}
\end{center}
\end{figure}
Fig.~\ref{fig:TwoBodyDist} displays event distributions as a function of $\cos\theta_X$ for two different scenarios. First, in blue, we show the expected spectrum for $PA_{\rm FB} = 1/2$ and 100 expected signal events. This is applicable for a perfectly-polarized DF $N$ decaying, for instance, to $\nu \pi^0$. In purple, we show the expected spectrum for $PA_{\rm FB} = -1/6$ and 150 expected signal events, which could be applicable for the decays $N \to \nu \gamma$, $N \to \nu \rho^0$, or $N \to \nu Z$, for certain choices of masses and couplings.

Our goal now is to determine how well the MF scenario can be excluded, given these two-body decays, for particular choices of masses and couplings, and assuming the DF hypothesis is true. In the MF scenario, the expected distribution is flat with respect to $\cos\theta_X$, while the data reflect the underlying linear relation of a DF.  We determine the best fit to the simulated data under the DF and MF scenarios and characterize the exclusion power with a log-likelihood ($\log\mathcal{L}$) approach.  It is instructive to first consider the situation where there is sufficient data that when binned in $n_b$ equal-sized bins in $\cos\theta_X$ the contents of each bin are normally distributed.  Then the log-likelihood follows a $\chi^2$ distribution and the $\Delta \chi^2$ between the MF hypothesis and the data is, on average,
\begin{equation}
\Delta\chi^2 = \frac{4}{3}N_{\rm evt.} \left(P A_{\rm FB}\right)^2 \left(\frac{n_b^2-1}{n_b^2}\right)~.
\end{equation}
Unsurprisingly, $\Delta\chi^2$ grows proportional to $N_{\rm evt.}$ and is quadratic with respect to $PA_{\rm FB}$ -- more events makes it easier to exclude the MF (flat with respect to $\cos\theta_X$) hypothesis, and the larger (in magnitude) the true $PA_{\rm FB}$, the easier this exclusion is as well.  With more limited statistics, i.e. the real world, the bin contents will be Poisson distributed and the expected $\Delta\chi^2$ is not so simple. This case we study numerically.  We maximize the likelihood for both the MF and DF hypotheses to explain the data and assume their difference follows a $\chi^2$ distribution, $\Delta \chi^2 \equiv -2\Delta \log\mathcal{L} = -2 \left(\log\mathcal{L}^{\rm max.}_{\rm DF} - \log\mathcal{L}^{\rm max.}_{\rm MF}\right)$.  

In practice, we also want to determine, for a given decay channel $N \to \nu X^0$, the number of signal events required to definitively rule out the MF hypothesis, i.e., $\Delta \chi^2 = 9$ ($3\sigma$ exclusion) or $25$ ($5\sigma$).
\begin{figure}
\begin{center}
\includegraphics[width=0.7\linewidth]{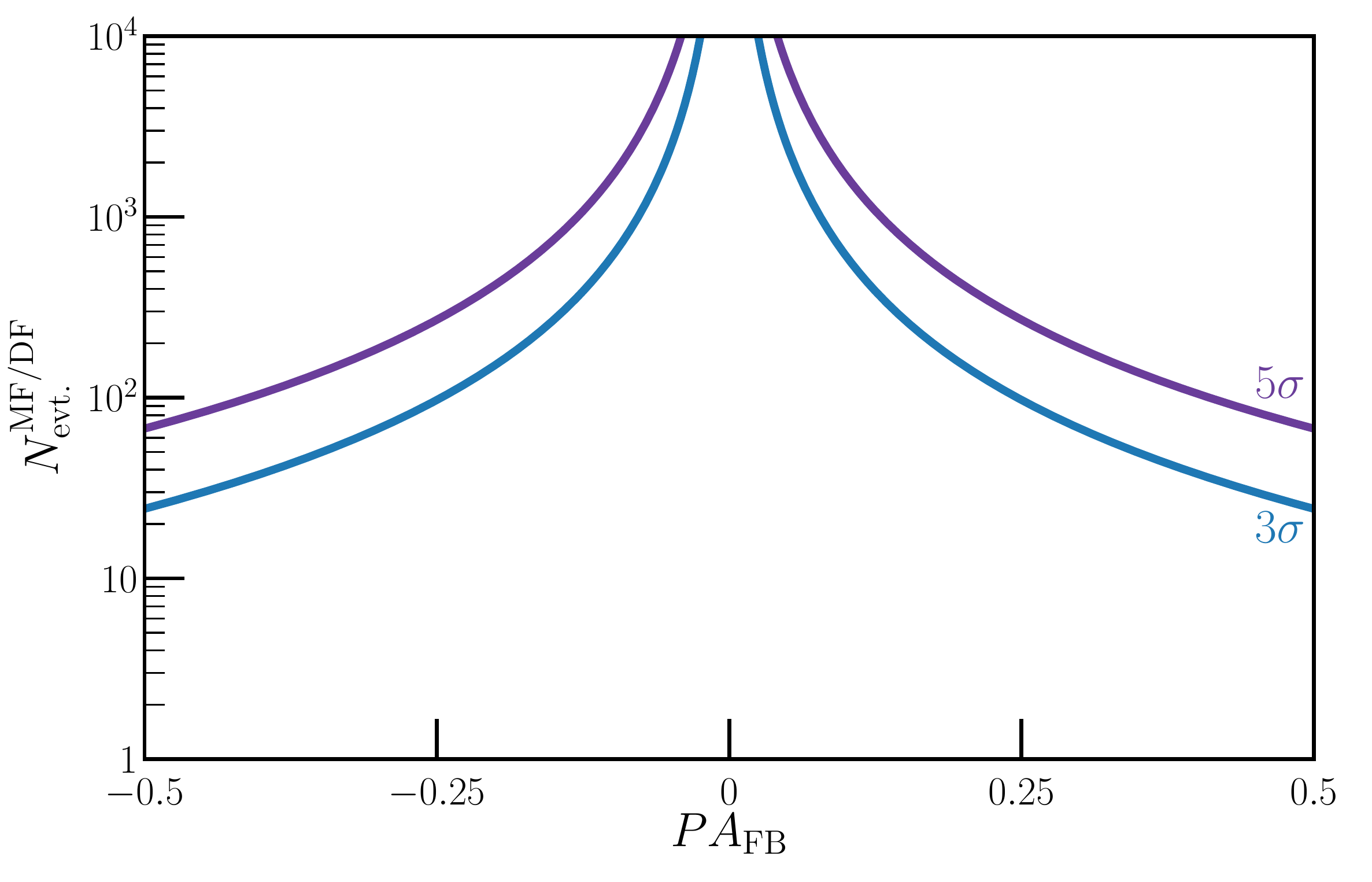}
\caption{Required number of expected true signal events (assuming $N$ is a Dirac fermion) to exclude the Majorana fermion hypothesis at $3\sigma$ confidence (blue) or $5\sigma$ confidence (purple).
\label{fig:TwoBodyExcl}}
\end{center}
\end{figure}
The number of expected signal events required to perform this analysis at $3\sigma$ ($5\sigma$) is shown as a blue (purple) line in Fig.~\ref{fig:TwoBodyExcl}, as a function of $PA_{\rm FB}$; this analysis was done using 20 bins, but the results are not expected to change significantly if the number of bins is changed. We see here that, even with a perfectly-polarized source of $N$, no backgrounds, and a maximal anistropy parameter, $\mathcal{O}(100)$ signal events are required to definitively rule out the MF hypothesis (at $5\sigma$ confidence). If the anisotropy is smaller -- for instance, if we are analyzing the decay $N \to \nu \rho^0$ and $m_N \approx 2m_\rho$, this expectation is $1/3$ -- then $\mathcal{O}(1000)$ events are required. If an upcoming experiment discovers a new particle $N$ in one of these decay channels with a handful of signal events, a significantly larger/higher-luminosity follow-up experiment will be required to identify whether $N$ is a DF or MF.

The scenario where $N$ decays to $\nu X^0$ and only one final-state kinematical variable $\cos\theta_X$ is relevant, is a simple one that can be understood with two input parameters (the expected number of signal events and the anisotropy parameter) and one-dimensional distributions like the ones shown in Fig.~\ref{fig:TwoBodyDist}. However, it serves as an illustrative example before exploring the more complicated scenarios we delve into in the following Sections~\ref{subsec:MFDF:YesSpin} and~\ref{subsec:MFDF:NoSpin}. These sections require more final-state kinematical variables and have a larger set of input parameters, and we will use more complicated statistical techniques in an attempt to separate the DF and MF hypotheses.

%%%%%%%%%%%%%%%%%%%%%%%%%%%%%%%%%%%%%%%%%%%%
\subsection{Three-Body Decays}\label{subsec:ThreeBodyDirMaj}
%%%%%%%%%%%%%%%%%%%%%%%%%%%%%%%%%%%%%%%%%%%%
For the remainder of this work, we focus on three-body decays of $N$, specifically $N \to \nu \ell_\alpha^- \ell_\beta^+$, where $\ell_\alpha$ and $\ell_\beta$ are charged leptons of flavors $\alpha$ and $\beta$, respectively. As discussed earlier, if $\alpha \neq \beta$, we expect that measuring the relative rates of the final states $\ell_\alpha^- \ell_\beta^+$ and $\ell_\alpha^+ \ell_\beta^-$ provides the strongest information for distinguishing between the MF and DF hypotheses. If $N$ is a MF, these rates must be equal, if $N$ is a DF, this is not guaranteed. Moving forward, we will focus on the case $\alpha = \beta$, where these counting experiments are not enough to separate the hypotheses.

We divide this discussion, and the remainder of this section, into two separate analyses. First, in Section~\ref{subsec:MFDF:YesSpin}, we assume that the $N$s are (at least partially) polarized. This allows us to use observables associated with its spin in our analysis. In doing so, we perform an extended likelihood analysis, described in Appendix~\ref{app:Unbinned}, to extract as much information as possible out of the kinematical observables. We will identify some cases in which the $N$s are unpolarized but the separation between the DF and MF hypotheses is still attainable. To further address this situation, in Section~\ref{subsec:MFDF:NoSpin}, we analyze the case where $N$ is unpolarized and therefore spin-related observables cannot be used for this distinction.

%%%%%%%%%%%%%%%%%%%%%%%%%%%%%%%%%%%%%%%%%%%%
\subsubsection{Observations with a Polarized Source}\label{subsec:MFDF:YesSpin}
%%%%%%%%%%%%%%%%%%%%%%%%%%%%%%%%%%%%%%%%%%%%
As motivated earlier, we will assume that $N$ is produced in the decay $K^+ \to \mu^+ N$ and that it decays via $N \to \nu e^+ e^-$, restricting its mass roughly to the range $[1, 385]$ MeV. Moreover, in our analysis we will assume that this production/decay channel is known and that $m_N$ is measured.

To determine how the complete set of observables assists in discriminating between the MF and DF hypotheses, we simulate data under the DF hypothesis for a given decay-interaction structure and determine how well the MF hypothesis can be excluded for these simulated data. We perform pseudoexperiments to determine the expected sensitivity (see Appendix~\ref{app:Unbinned}). The key quantity we determine using this procedure is the number of expected events that is required to reject the MF hypothesis at $3\sigma$ confidence.

We explore three test cases here, allowing the true value of $m_N$ to vary in each. The three are
\begin{description}
\item[Case 1] Pure scalar/pseudoscalar interactions with maximal forward/backward asymmetry: $G_{SS} = G_{SP} = G_{PS} = G_{PP}$. 
\item[Case 2] Vector/axial-vector interactions $G_{VV} = -G_{AV} = -G_{VA} = G_{AA}$. This is similar to the scenario in which $N$ is a heavy neutral lepton mixing predominantly with $\nu_e$. 
\item[Case 3]  Vector/axial-vector interactions $G_{VV} = G_{AV} = 0$ and $G_{VA} = -G_{AA}$. This is similar to the scenario in which $N$ is a heavy neutral lepton mixing predominantly with $\nu_\mu$ and/or $\nu_\tau$. 
\end{description}
These were identified as interesting cases in Ref.~\cite{deGouvea:2021ual} for a variety of reasons. Case 1 provides maximal forward/backward asymmetry which is difficult for a MF decay distribution to mimic. However, in the absence of polarization, the MF/DF distinction vanishes and it should be impossible to rule out the MF hypotheses. Cases 2 and 3, in addition to their relevance to many phenomenological studies, yield some non-zero forward/backward asymmetry, and the $\nu_e$-like mixing scenario (Case 2) has some MF/DF distinction even without polarization, whereas the $\nu_{\mu,\tau}$-like mixing scenario (Case 3) does not (we will see Cases 2 and 3 map onto specific cases of interest in Section~\ref{subsec:MFDF:NoSpin} as long as $m_N \gg m_e$). Cases 2 and 3 are similar to various HNL-like scenarios. In particular, Case 2 is what is expected if the HNL mixes only with $\nu_e$ and the weak mixing angle were 0, and Case 3 corresponds to no $\nu_e$ mixing and $\sin^2\theta_w = 1/4$. The case of an HNL with arbitrary mixing and with the weak mixing angle at its physical value will be discussed in more detail in Section~\ref{subsec:HNL_DiracVsMajorana}.

\begin{figure}
\begin{center}
\includegraphics[width=0.7\linewidth]{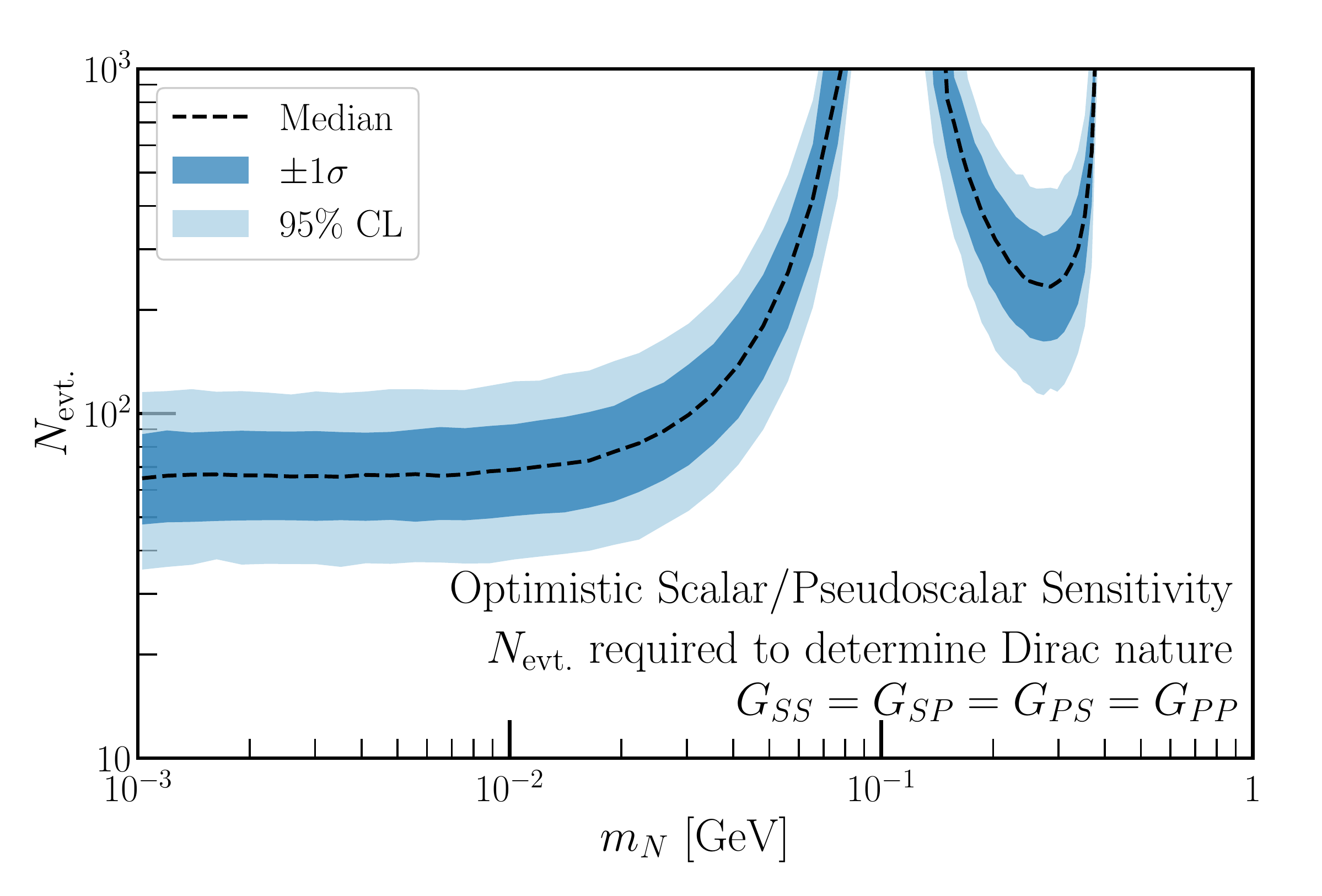}
\caption{Number of events as a function of the $N$ mass $m_N$ required to rule out the Majorana-fermion hypothesis at 3$\sigma$ confidence assuming that the Dirac Fermion hypothesis, with $G_{SS} = G_{SP} = G_{PS} = G_{PP}$, is true. See Appendix~\ref{app:Unbinned} for explanation of statistical techniques. The dashed black line corresponds to the median expected capability, whereas the dark (light) blue filled regions show the $\pm 1\sigma$ ($95\%$) expectation.
\label{fig:MFDF:YesSpin:SP}}
\end{center}
\end{figure}

First, let us determine the number of events required to reject the MF hypothesis if $N$ is a DF with the characteristics of Case 1. 
%For $m_N$ between roughly $1$ MeV and $385$ MeV, we simulate pseudoexperiments according to the procedure described in Appendix~\ref{app:Unbinned}. 
We determine the expected distribution of the preference for the null hypothesis (DF, any coupling allowed) over the alternative one (MF, any coupling allowed). The median expectation, translated into how many events are required to prefer the null hypothesis at high confidence, is shown as a black dashed line in Fig.~\ref{fig:MFDF:YesSpin:SP}. As expected, the distinction becomes more difficult (i.e., requires more events) when $N$ is less polarized -- both when $m_N \approx m_\mu$, as well as when $m_N$ is between $m_\mu$ and $m_{K^\pm} - m_\mu$, see Fig.~\ref{fig:AlphaPolDir}.
When performing pseudoexperiments, we gain access to the \textit{distribution} of expected preference of one hypothesis over another. This allows us to also determine how many events are required to reject the MF hypothesis if statistical fluctuations cause us to be lucky or unlucky. The band of this expectation at $\pm 1\sigma$ ($\pm 2\sigma$) is shown as a filled-in dark blue (light blue) region in Fig.~\ref{fig:MFDF:YesSpin:SP}. Within $1\sigma$ expectation, we could require fewer events (${\sim}50$ as opposed to ${\sim}60$) for a wide range of masses. Similarly, more events (${\sim}90$) may be required if a fluctuation is in the other direction.

\begin{figure}
\begin{center}
\includegraphics[width=\linewidth]{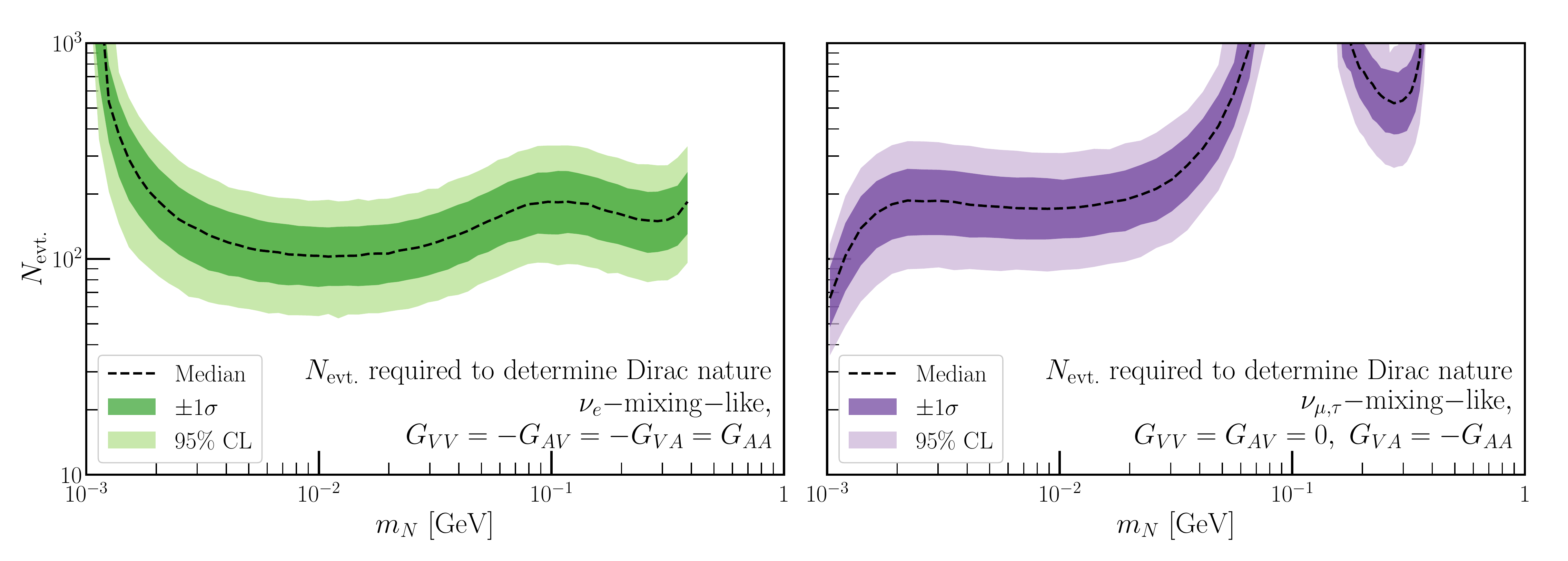}
\caption{Number of events as a function of the $N$ mass $m_N$ required to rule out the Majorana-fermion hypothesis at $3\sigma$ confidence assuming that the Dirac Fermion hypothesis is true. Two different assumptions are made about the true distributions, corresponding to the two vector/axial-vector cases discussed in the text and as labelled on the respective figures. See Appendix~\ref{app:Unbinned} for explanation of statistical techniques. The dashed black line corresponds to the median expected capability, whereas the dark (light) filled regions show the $\pm 1\sigma$ ($\pm 95\%$) expectation.
\label{fig:MFDF:YesSpin:VA}}
\end{center}
\end{figure}

We repeat this process for the Vector/Axial-vector cases in Fig.~\ref{fig:MFDF:YesSpin:VA}, where the left (right) panel corresponds to the $\nu_e$- ($\nu_{\mu,\tau}$-)mixing-like case. Different colors are used only for clarity here. As with Fig.~\ref{fig:MFDF:YesSpin:SP}, the dark (light) filled regions correspond to $\pm 1\sigma$ ($95\%$) expectations, and the dashed black lines correspond to the median expectation.
In general, for the vector/axial-vector cases in Fig.~\ref{fig:MFDF:YesSpin:VA}, we note that more events are required to reject the MF hypothesis at high confidence -- $\mathcal{O}(100-200)$ generally -- than when the true hypothesis was scalar/pseudoscalar interactions (in Fig.~\ref{fig:MFDF:YesSpin:SP}). This arises from the fact that the scalar/pseudoscalar decay distributions of a DF $N$ have more forward-backward asymmetry, a signature that enables separation between the MF and DF hypotheses.

One additional feature of Fig.~\ref{fig:MFDF:YesSpin:VA} is particularly notable -- for $m_N \approx m_\mu$, the $N$ source will arrive at the detector unpolarized, yet we see in the $\nu_e$-mixing-like case (green, left panel) that MF/DF separation is still possible given enough events. This is not the case for the $\nu_{\mu,\tau}$-mixing-like case (purple, right panel), where the MF/DF separation becomes impossible at this mass. This implies that, even without a polarized source (or, equivalently, when the distributions' dependence on $\cos\theta_{\ell\ell}$ and $\gamma_{\ell\ell}$ vanish) we can still potentially determine whether $N$ is a DF or MF. 

%%%%%%%%%%%%%%%%%%%%%%%%%%%%%%%%%%%%%%%%%%%%
\subsubsection{Observations with an Unpolarized Source}\label{subsec:MFDF:NoSpin}
%%%%%%%%%%%%%%%%%%%%%%%%%%%%%%%%%%%%%%%%%%%%
As noted above, it is possible that the $N$ source at the detector is unpolarized, for instance, when $m_N \approx m_{\ell'}$ (where $m_{\ell'}$ is the charged lepton with which the $N$ source is produced, e.g., the muon in $K^+ \to \mu^+ N$). Following the argument around Eq.~\eqref{eq:dGammaPol}, we see that if $P \to 0$, all spin-dependent terms of the matrix-element-squared will no longer contribute to the decay distribution of $N$. This implies that the distribution is flat with respect to $\cos\theta_{\ell\ell}$ and $\gamma_{\ell\ell}$ regardless of whether $N$ is a DF or MF, and we are left with only two relevant kinematical variables.  The two relevant variables are $z_{\ell\ell} \equiv m_{\ell\ell}^2/m_N^2$ and $z_{\nu m} \equiv m_{\nu m}^2/m_N^2$.
 For further simplicity, we consider the scenario in which $N$ is significantly more massive than the charged leptons into which it is decaying, i.e., $m_{\ell}/m_N \approx 0$. If we are interested in the regime where $m_N \approx m_\mu$ (and $N$ is coming from $K^+ \to \mu^+ N$ decay) followed by subsequent $N \to \nu e^+ e^-$ decay, this assumption will hold.

When $P \approx 0$, we are only sensitive to spin-independent terms in the $N$ decay matrix-element-squared, i.e. the six Lorentz Invariants $K_{j}$; $j \in [1, 6]$~\cite{deGouvea:2021ual}. Moreover, in the limit $m_N \gg m_{e}$, only three Lorentz invariants are relevant, $K_4$, $K_5$, and $K_6$ (which have respective coefficients $C_4$, $C_5$, and $C_6$). Given that these are the only three relevant $C_{j}$, and that many linear combinations appear repeatedly, we perform some redefinitions:
\begin{comment}
The three important coefficients for this scenario under the DF hypothesis are~\cite{deGouvea:2021ual}
\begin{align}
C_4^\mathrm{DF} &= 16\left[ \left\lvert G_{AA} + G_{VV}\right\vert^2 + \left\lvert G_{VA} + G_{AV}\right\rvert^2\right] + 128 \left \lvert G_{TT}\right\rvert^2 - 32\mathrm{Re}\left[ \left(G_{SS} + G_{PP}\right) G_{TT}^*\right], \\
C_5^\mathrm{DF} &= 16\left[ \left\lvert G_{AA} - G_{VV}\right\vert^2 + \left\lvert G_{VA} - G_{AV}\right\rvert^2\right] + 128 \left \lvert G_{TT}\right\rvert^2 + 32\mathrm{Re}\left[ \left(G_{SS} + G_{PP}\right) G_{TT}^*\right], \\
C_6^\mathrm{DF} &= 8\left[ \left\lvert G_{PP}\right\rvert^2 + \left\lvert G_{SP}\right\rvert^2 + \left\lvert G_{PS}\right\rvert^2 + \left\lvert G_{SS}\right\rvert^2\right] - 64 \left\lvert G_{TT}\right\rvert^2.
\end{align}
Since many linear combinations appear repeatedly, we perform some redefinitions:
\end{comment}
\begin{equation}\label{eq:CsNoPolDir}
\left( \begin{array}{c} C_4 \\ C_5 \\ C_6 \end{array}\right)^\mathrm{DF} = \displaystyle\left(\begin{array}{c} \displaystyle V\cos^2\beta + \frac{2T}{3} + \sqrt{\frac{2ST}{3}}\cos\theta_I \\[8pt] \displaystyle V\sin^2\beta + \frac{2T}{3} - \sqrt{\frac{2ST}{3}}\cos\theta_I \\[8pt] \displaystyle S - \frac{T}{3}\end{array}\right) ~,
\end{equation}
with $S,V,T\ge 0$. These $S$, $V$, and $T$ are purely functions of $G_{NL}$ and $\overline{G}_{NL}$, introduced in Eqs.~\eqref{eq:MGNL} and~\eqref{eq:MbGNL}. Similarly, for the MF hypothesis
\begin{equation}\label{eq:CsNoPolMaj}
\left( \begin{array}{c} C_4 \\ C_5 \\ C_6 \end{array}\right)^\mathrm{MF} = 
 \displaystyle\left(\begin{array}{c} \displaystyle\frac{V}{2} + \frac{2T}{3} \\[8pt] \displaystyle\frac{V}{2} + \frac{2T}{3}  \\[8pt] \displaystyle S - \frac{T}{3}\end{array}\right)~.
\end{equation}
The total width in either the DF or MF scenario is $\propto (S + V + T)$.

From the forms of Eqs.~\eqref{eq:CsNoPolDir},\eqref{eq:CsNoPolMaj}, it is clear there are certain scenarios that can not be distinguished with an unpolarized source alone. For instance, if both $\cos\theta_I=0$ and $\tan\beta=1$ the two hypotheses are indistinguishable, if the only nonzero value from $\left\lbrace S, V, T\right\rbrace$ is $S$ \textbf{\textit{or}} $T$ then no distinction is possible, and if $V$ is the only nonzero among $\left\lbrace S, V, T\right\rbrace$ and $\tan\beta = 1$  then the distinction is impossible.  We ignore these situations from now on.  Instead, we consider situations in which the two hypotheses predict different distributions and determine how much data is necessary to tell them apart.

We simulate events %distributed as a function of $z_{\ell\ell}$ and $z_{\nu m}$ depending on 
for different truth assumptions of $\left\lbrace S,~V,~T,~\tan\beta,~\cos\theta_I\right\rbrace_\mathrm{true}$ and then perform fits assuming either the DF or MF hypotheses. The quantity of interest is the difference in the log-likelihood between the best-fit points according to either of these hypotheses, $-2\Delta \log\mathcal{L}$. Since the DF hypothesis contains two more free parameters ($\tan\beta$ and $\cos\theta_I$) than the MF one, and we assume this test statistic follows a $\chi^2$-distribution, we use $\Delta\chi^2 = 11.83$ ($28.74$) as the threshold for $3\sigma$ ($5\sigma$) rejection of the MF hypothesis.

\begin{figure}
\begin{center}
\includegraphics[width=0.65\linewidth]{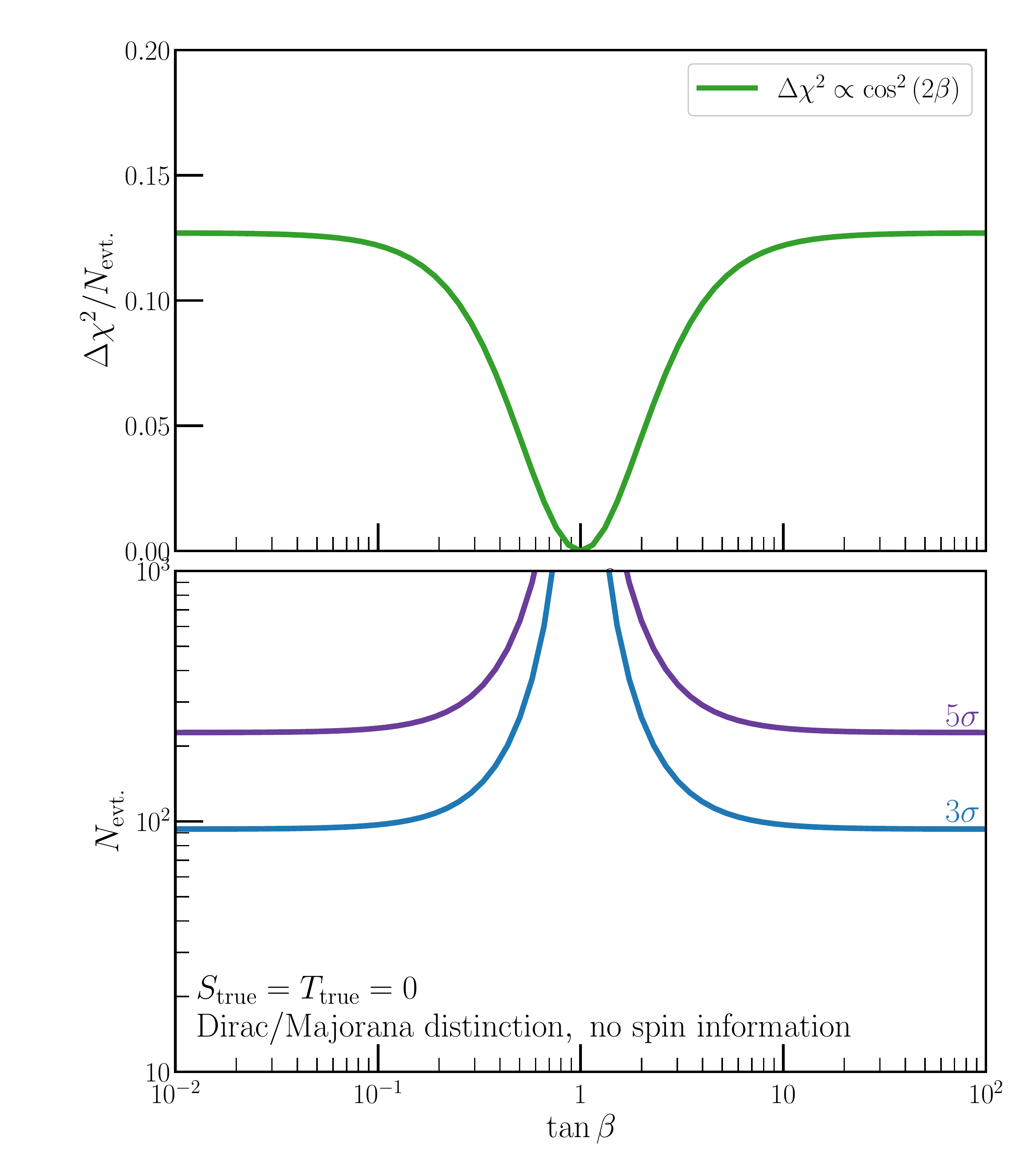}
\caption{Top: dependence of the test statistic $\Delta \chi^2$ to separate the Dirac-fermion and Majorana-fermion hypotheses in the scenario in which $V_\mathrm{true}$ is the only nonzero contribution to the Dirac fermion signal, with the truth parameter $\tan\beta$ varied as well. Bottom: the number of signal events $N_{\rm evt.}$ required for $3\sigma$ (blue) or $5\sigma$ (purple) rejection of the Majorana fermion hypothesis as a function of $\tan\beta$. \label{fig:DM:VNoSpin}}
\end{center}
\end{figure}

First consider the situation where $V_\mathrm{true}$ is the only nonzero contribution, $S_\mathrm{true} = T_\mathrm{true} = 0$.  In the limit of large statistics, with $N_{\rm evt.}$ the total number of events seen\footnote{We assume that we are in the long-lifetime limit where $N_{\rm evt.} = \mathcal{L} \Gamma$, where $\mathcal{L}$ includes the experimental configuration such as number of protons delivered, detector volume, etc., and $\Gamma\ \propto\ S_{\rm true} + V_{\rm true} + T_{\rm true}$.}, the difference between the MF and DF hypotheses evaluated at the same parameter point is
\begin{equation}
\Delta \chi^2 = \left(\frac{6C-5}{4}\right) N_{\rm evt.} \cos^2 2\beta \approx 0.124\, N_{\rm evt.}\cos^2 2\beta~,
\end{equation}
where $C\approx 0.916$ is Catalan's constant.  As before, for more limited statistics, a numerical determination of the discriminatory power is necessary; nonetheless, we expect the scaling with $N_{\rm evt.}$ and $\beta$ to remain the same.  Fig.~\ref{fig:DM:VNoSpin}(top) presents the behavior of $\Delta\chi^2$ as a function of $\tan\beta$, demonstrating that the MF/DF separation is maximized for $\tan\beta \to 0$ or $\infty$ (as discussed above) and, without any spin information, this separation is impossible for $\tan\beta = 1$. The bottom panel of Fig.~\ref{fig:DM:VNoSpin} displays the number of expected signal events $N_{\rm evt.}$ required for $3\sigma$ (blue) or $5\sigma$ (purple) rejection of the MF hypothesis as a function of the true $\tan\beta$ -- at best, $3\sigma$ discrimination requires $\mathcal{O}(100)$ signal events if no spin/polarization information can be utilized. In contrast, depending on the $N$ source polarization, $m_N$ and the specific couplings assumed, we can expect to differentiate the DF and MF hypotheses at high confidence with $\mathcal{O}(90)$ events, see Fig.~\ref{fig:MFDF:YesSpin:VA}.

\begin{figure}
\begin{center}
\includegraphics[width=0.7\linewidth]{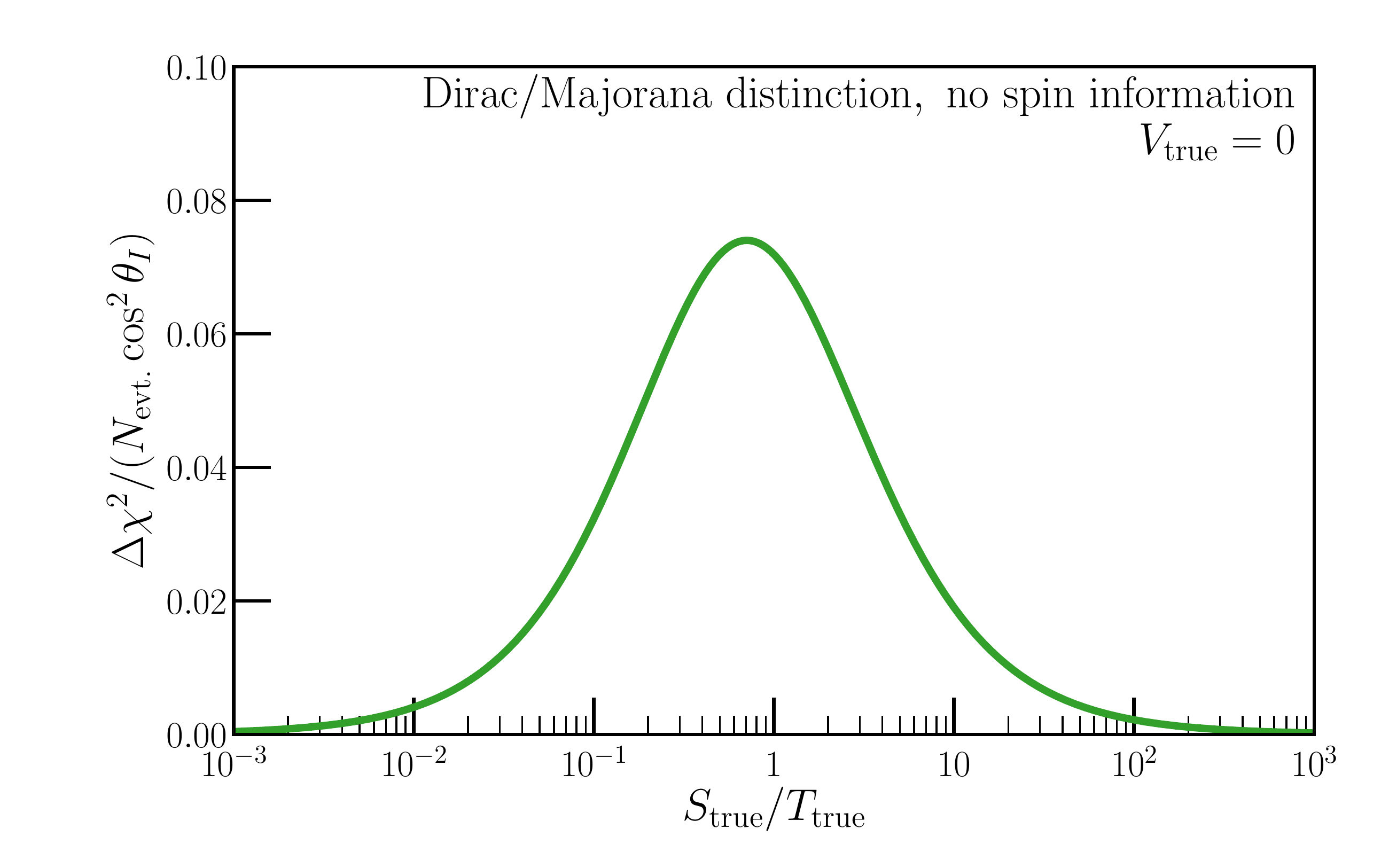}
\caption{Behavior of the $\Delta \chi^2$ to separate the Dirac fermion and Majorana fermion hypotheses (normalized to the number of expected events and $\cos^2\theta_I$) in the scenario where $V_{\rm true}$ is zero. The $\Delta \chi^2$ function depends on this ratio as explained in the text, see Eq.~\eqref{eq:DChiST}. \label{fig:DM:SPTNoSpin}}
\end{center}
\end{figure}

The second scenario we consider is with nonzero $S_\mathrm{true}$ and $T_\mathrm{true}$, fixing $V_\mathrm{true} = 0$. As discussed above, if either $S$ or $T$ is zero, then $C_4 = C_5$ and the DF/MF distinction is impossible.  In the limit of large statistics the test statistic is
\begin{equation}\label{eq:DChiST}
\Delta\chi^2 = \frac{N_{\rm evt.} S_\mathrm{true}}{S_\mathrm{true}+T_\mathrm{true}} \cos^2\theta_I \times \zeta\left(\frac{S_\mathrm{true}}{T_\mathrm{true}}\right),
\end{equation}
$\zeta(S/T)$ is a complicated function that depends only on the relative ratio of these two parameters.  It has simple limits $\zeta(x\rightarrow 0) =2(64\log 2-37)/35$  and $\zeta(x\rightarrow \infty) = 2/(9x)$. The full form of $\Delta\chi^2$ in this case, divided by $N_{\rm evt.}\cos^2\theta_I$, is presented in Fig.~\ref{fig:DM:SPTNoSpin} and is maximized when $S_\mathrm{true} \approx 0.7\, T_\mathrm{true}$ and goes to zero at either extreme.
The rejection of the MF hypothesis is easiest when $\cos^2\theta_I = 1$ and when this ratio between $S$ and $T$ is $\mathcal{O}(1)$ -- even in such an optimistic scenario, the required number of signal events is large -- $\mathcal{O}(200)$ for $3\sigma$ rejection and $\mathcal{O}(400)$ for $5\sigma$ rejection. This is to be compared with the result of Fig.~\ref{fig:MFDF:YesSpin:SP}, where we saw that, when the $N$s are polarized, this distinction can require as few as $\mathcal{O}(50)$ events.

%%%%%%%%%%%%%%%%%%%%%%%%%%%%%%%%%%%%%%%%%%%%%%%%
\subsection{Dirac/Majorana Fermion Distinction for Heavy Neutral Leptons}
\label{subsec:HNL_DiracVsMajorana}
%%%%%%%%%%%%%%%%%%%%%%%%%%%%%%%%%%%%%%%%%%%%%%%%
The final case we wish to consider is when the newly-discovered fermion is assumed to only interact through the exchange of $W$ and $Z$ bosons, via mass mixing of $N$ with the light neutrinos. In this case, the operators that generate three-body decay matrix elements are related to the weak interactions of the SM.
%, all proportional to $G_F$; Ref.~\cite{deGouvea:2021ual} discussed this case in some detail. 
Consider initially the spin-independent $C_i$, which would be the only available information for an unpolarized source of HNLs. Under this assumption, $S = T = 0$ and only $V$ is nonzero. We can determine the predicted value of $\tan\beta$ in Eq.~\eqref{eq:CsNoPolDir} under the DF hypothesis. Still focusing on the decay channel $N \to \nu e^+ e^-$, the contributions from electron mixing ($U_{eN}$) provide both charged- and neutral-current decays, whereas contributions from muon/tau mixing ($U_{\mu N}$ and $U_{\tau N}$) provide only neutral-current ones. Thus,
\begin{equation}
\tan^2\beta = \frac{4 s_w^4 \left( \absq{U_{eN}} + \absq{U_{\mu N}} + \absq{U_{\tau N}}\right)}{\absq{U_{eN}} \left(1 + 2s_w^2\right)^2 + \left( \absq{U_{\mu N}} + \absq{U_{\tau N}} \right)\left(1 - 2 s_w^2\right)^2},
\end{equation}
where $s_w^2 \approx 0.223$ is the sine-squared of the weak mixing angle. The value of $\tan\beta$ depends on the ratio $(\absq{U_{\mu N}} + \absq{U_{\tau N}})/\absq{U_{eN}}$. In the limit where $|U_{eN}| \gg |U_{\mu,\tau N}|$, $\tan\beta \to 0.308$. In the opposite regime, $|U_{\mu,\tau N}| \gg |U_{eN}|$, $\tan\beta \to 0.805$.  Both of these scenarios can, in principle, be distinguished from the MF prediction where $\tan\beta = 1$. 

Going beyond the spin-independent $C_i$, the DF scenario also predicts a non-zero forward/backward asymmetry in the decay distribution of $N \to \nu e^+ e^-$. Regardless of whether the decay proceeds via a combination of neutral and charged currents or only neutral ones, we find that $A_{\rm FB} = 1/6 + \mathcal{O}(m_e^2/m_N^2)$. This is in contrast with the MF prediction of a forward/backward symmetric distribution, $A_{\rm FB} = 0$.

\begin{figure}
\begin{center}
\includegraphics[width=0.8\linewidth]{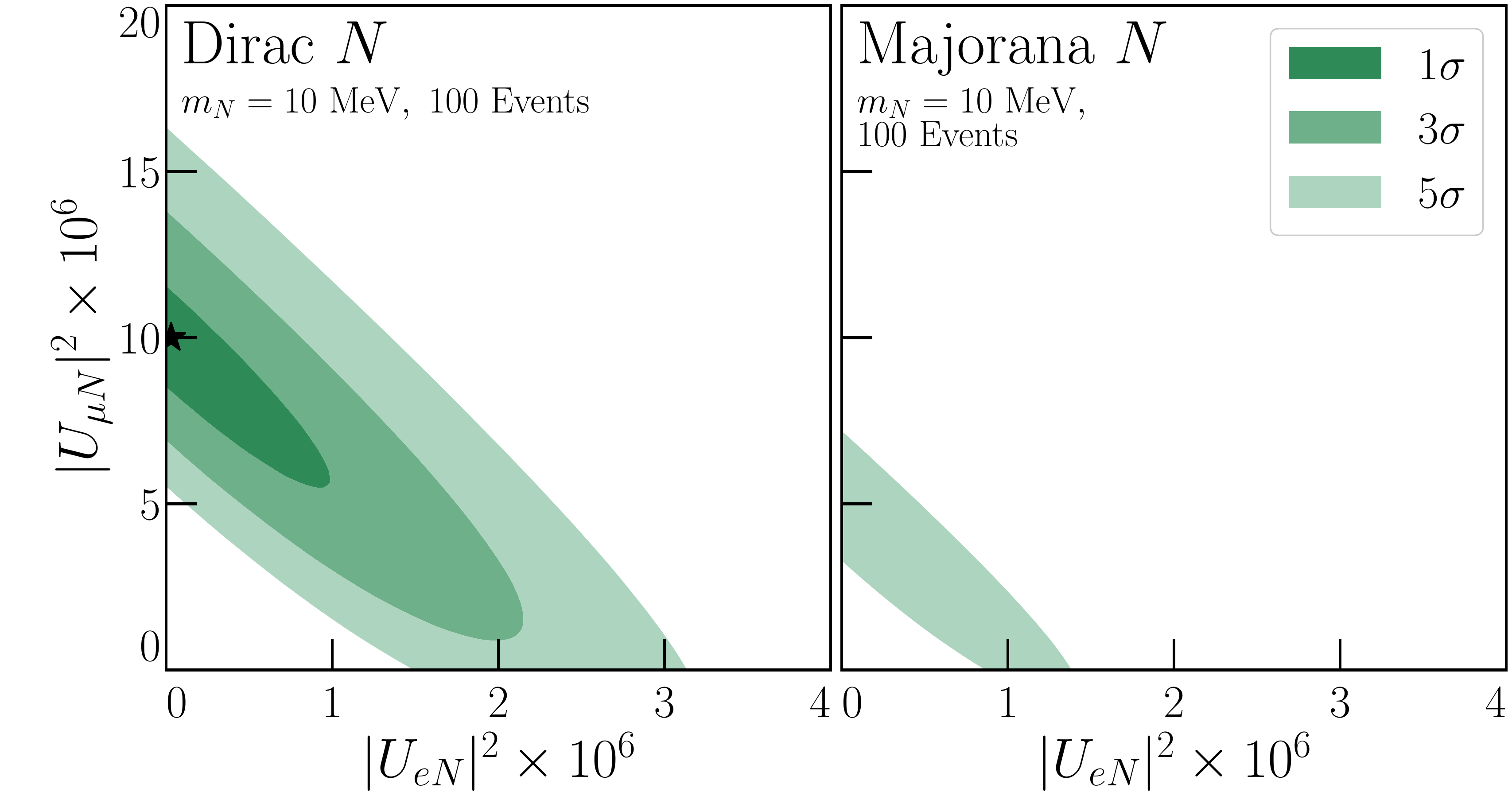}
\caption{Expected measurement capability at $1$, $3$, and $5\sigma$ (dark, medium, and light green regions, respectively) when assuming that a $10$ MeV Dirac fermion is decaying purely via muon-mixing, i.e. $\left\lvert U_{\mu N}\right\rvert^2$ is the only non-zero mixing angle. We assume that 100 signal events are reconstructed, and determine regions of parameter space consistent with simulated data at the displayed confidence levels. When $N$ is assumed to be a Dirac fermion, but fit under the Majorana fermion hypothesis (right panel), no parameter space at $1$ or $3\sigma$ compatibility is found. \label{fig:HNLUSq}}
\end{center}
\end{figure}

In order to quantify how well-separated these hypotheses are, and how well certain parameters of the model may be measured, we perform pseudoexperiments. We assume that a DF HNL $N$ exists with mass $10$ MeV, and that its mixing with the SM neutrinos is via only a nonzero $\absq{U_{\mu N}} = 10^{-5}$ and that this mixing corresponds to an expectation of $100$ signal events in our next-generation experiment -- current experiments constrain $\absq{U_{\mu N}} \lesssim 2 \times 10^{-5}$~\cite{PiENu:2015seu,deGouvea:2015ndi,Bolton:2019pcu} and current/next-generation experiments have discovery potential at the level of $\mathcal{O}(10^{-5})$~\cite{Kelly:2021xbv,Arguelles:2021dqn} for similar $m_N$. We perform fits to the pseudodata under the DF and MF HNL hypotheses and determine the regions of parameter space where our expected measurements would lie at $1$, $3$, and $5\sigma$ confidence. This is presented in Fig.~\ref{fig:HNLUSq}. We see that the MF HNL hypothesis can be rejected in this case at over $3\sigma$ confidence. With fewer events, these expectations would be naturally less confident. Under the MF hypothesis, the best-fit point is at $\absq{U_{e N}} = 0$, $\absq{U_{\mu N}} = 5 \times 10^{-6}$, because a MF with a given mixing predicts a factor of two larger decay width (and therefore event rate) than a DF with the same mixing.

Restricting ourselves to the DF hypothesis (the left panel of Fig.~\ref{fig:HNLUSq}), we also see that 100 signal events is sufficient to rule out the electron-mixing-only hypothesis at just over $3\sigma$ confidence, due to the difference induced by the charged-current contribution if $\absq{U_{e N}} \neq 0$ vs. the neutral-current-only decay if only $\absq{U_{\mu N}}$ is nonzero. Note that, to explain the data, the elecron-mixing-only scenario prefers a smaller value of $\absq{U_{e N}} \approx 2 \times 10^{-6}$ compared to the true value of $\absq{U_{\mu N}} = 10^{-5}$. This is because, in the electron-mixing-only scenario, the partial with of $N \to \nu e^+ e^- \propto \absq{U_{e N}} \left( 1 + 4 s_w^2 + 8 s_w^4\right) \approx 2.29 \absq{U_{e N}}$ compared to the muon-mixing-only scenario where $\Gamma(N\to \nu_\mu e^+ e^-) \propto \absq{U_{\mu N}} \left( 1 - 4 s_w^2 + 8 s_w^4\right) \approx 0.51 \absq{U_{\mu N}}$ -- a significantly larger $\absq{U_{\mu N}}$ is required to have the same partial width compared to $\absq{U_{e N}}$.

%%%%%%%%%%%%%%%%%%%%%%%%%%%%%%%%%%%%%%%%%%%%
\section{Identifying the Coupling Structure if the Fermion Nature is Known}
\label{sec:InteractionStructure}
%%%%%%%%%%%%%%%%%%%%%%%%%%%%%%%%%%%%%%%%%%%%
Beside the question of whether a newly-detected $N$ is a DF or MF, another obvious question is to determine the nature of the interactions responsible for a particular decay mode. As discussed throughout this work and in Ref.~\cite{deGouvea:2021ual}, the contributions of different interaction types yield qualitatively different decay distributions.

Here, we assume, unless otherwise noted,  that enough information has been collected that the nature of $N$, i.e., whether it is a DF or MF, is known. We will generate pseudodata assuming a particular interaction structure and then address how well alternative interaction-structure hypotheses can be ruled out. The format of this section follows that of Section~\ref{subsec:ThreeBodyDirMaj}:  in Section~\ref{subsec:IntStr:YesSpin} we discuss how this differentiation can be performed if the $N$s are polarized and in Section~\ref{subsec:IntStr:NoSpin}, for cases where polarization is expected to be zero, we explore the possibilities that still remain.

%%%%%%%%%%%%%%%%%%%%%%%%%%%%%%%%%%%%%%%%%%%%
\subsection{Observations with a Polarized Source}\label{subsec:IntStr:YesSpin}
\begin{figure}
\begin{center}
\includegraphics[width=0.8\linewidth]{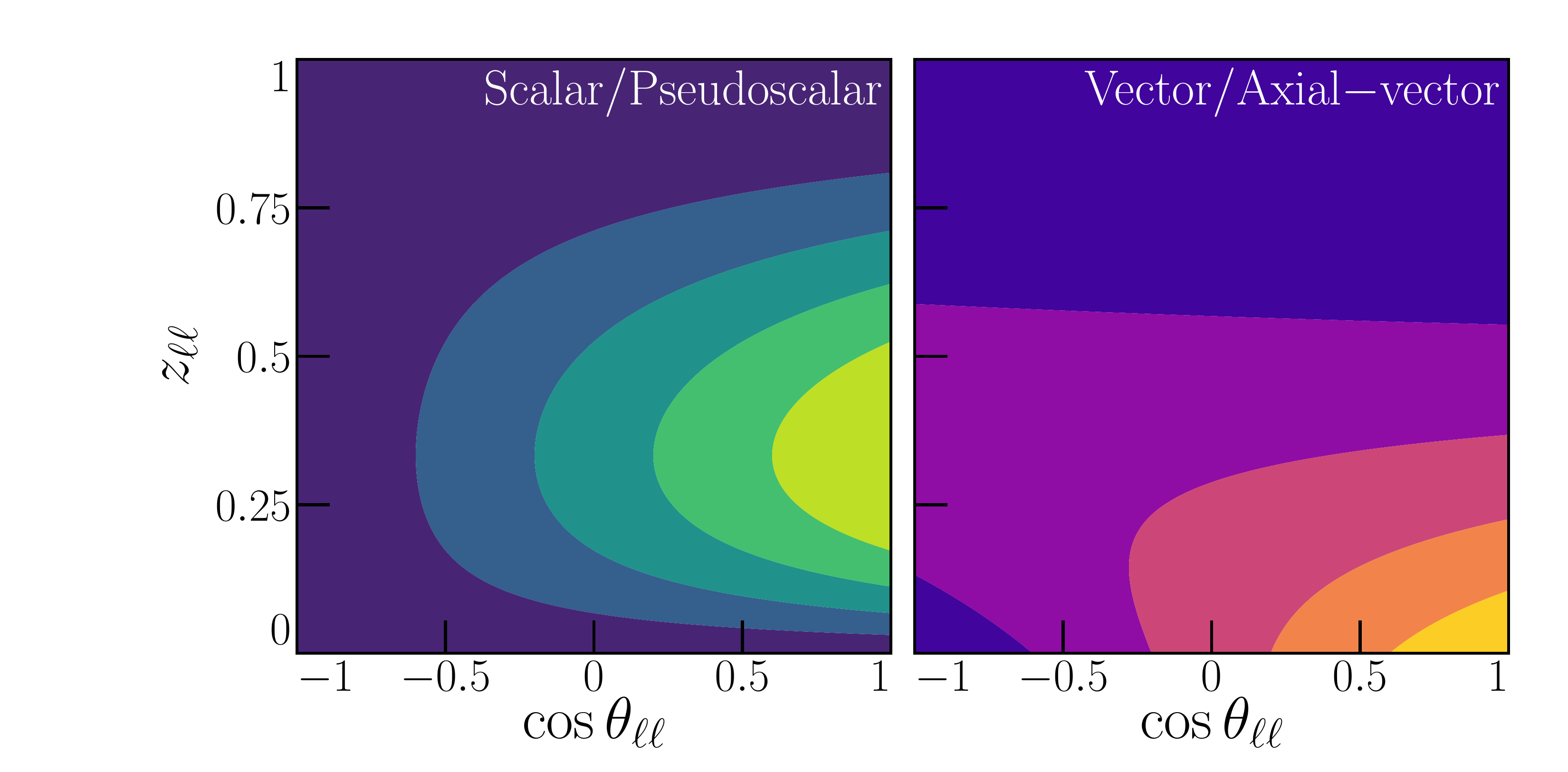}
\caption{Two-dimensional decay distributions as a function of the reduced charged-lepton invariant mass $z_{\ell\ell}$ vs. the charged-lepton pair direction $\theta_{\ell\ell}$ assuming $m_N \gg m_e$ for either scalar/pseudoscalar interactions (left) and vector/axial-vector interactions (right).
\label{fig:SVSpinInformation}}
\end{center}
\end{figure}

%%%%%%%%%%%%%%%%%%%%%%%%%%%%%%%%%%%%%%%%%%%%
If we have a source of polarized $N$, we can account for spin information in our analysis. Ref.~\cite{deGouvea:2021ual} explored how the angular dependence of the decay distribution of $N$ depends on different contributions to the matrix-element-squared. To illustrate this effect, we display the differential decay distribution for a subset of final-state kinematical observables, under different interaction hypotheses, in Fig.~\ref{fig:SVSpinInformation}. We show the dependence on $z_{\ell\ell}$ and $\cos\theta_{\ell\ell}$. Here, we assume that $N$ is a DF, that $m_N \gg m_e$, and that the $N$ is $100\%$ polarized. Lighter (darker) regions in each color scheme correspond to regions of parameter space where more (fewer) events are expected. 

Generically, the two-dimensional distribution for the scalar/pseudoscalar and vector/axial-vector cases over these two observables is
\begin{subnumcases}
{\label{eq:DiracAllCases}
\dfrac{1}{\Gamma}\dfrac{d\Gamma}{dz_{\ell\ell} d\cos\theta_{\ell\ell}} =}
6\left(1 - z_{\ell\ell}\right)^2 z_{\ell\ell} \left( 1 + \cos\theta_{\ell\ell}\right), & Scalar/Pseudoscalar, \label{eq:SP2D}\\
\left( 1 - z_{\ell\ell}\right)^2 \left( 1 + 2z_{\ell\ell} + \left(1 - 2z_{\ell\ell}\right)\cos\theta_{\ell\ell}\right), & Vector/Axial-vector. \label{eq:VA2D}
\end{subnumcases}
Notice that in the scalar/pseudoscalar case, Eq.~\eqref{eq:SP2D}, the ratio of the $\theta_{\ell\ell}$-dependent term and the $\theta_{\ell\ell}$-independent term is independent of $z_{\ell\ell}$, whereas for the vector/axial-vector case, Eq.~\eqref{eq:VA2D}, the ratio changes sign at $z_{\ell\ell} = 1/2$. This feature allows for separation of the scalar/pseudoscalar and vector/axial-vector decay hypotheses.

\begin{figure}
\begin{center}
\includegraphics[width=0.6\linewidth]{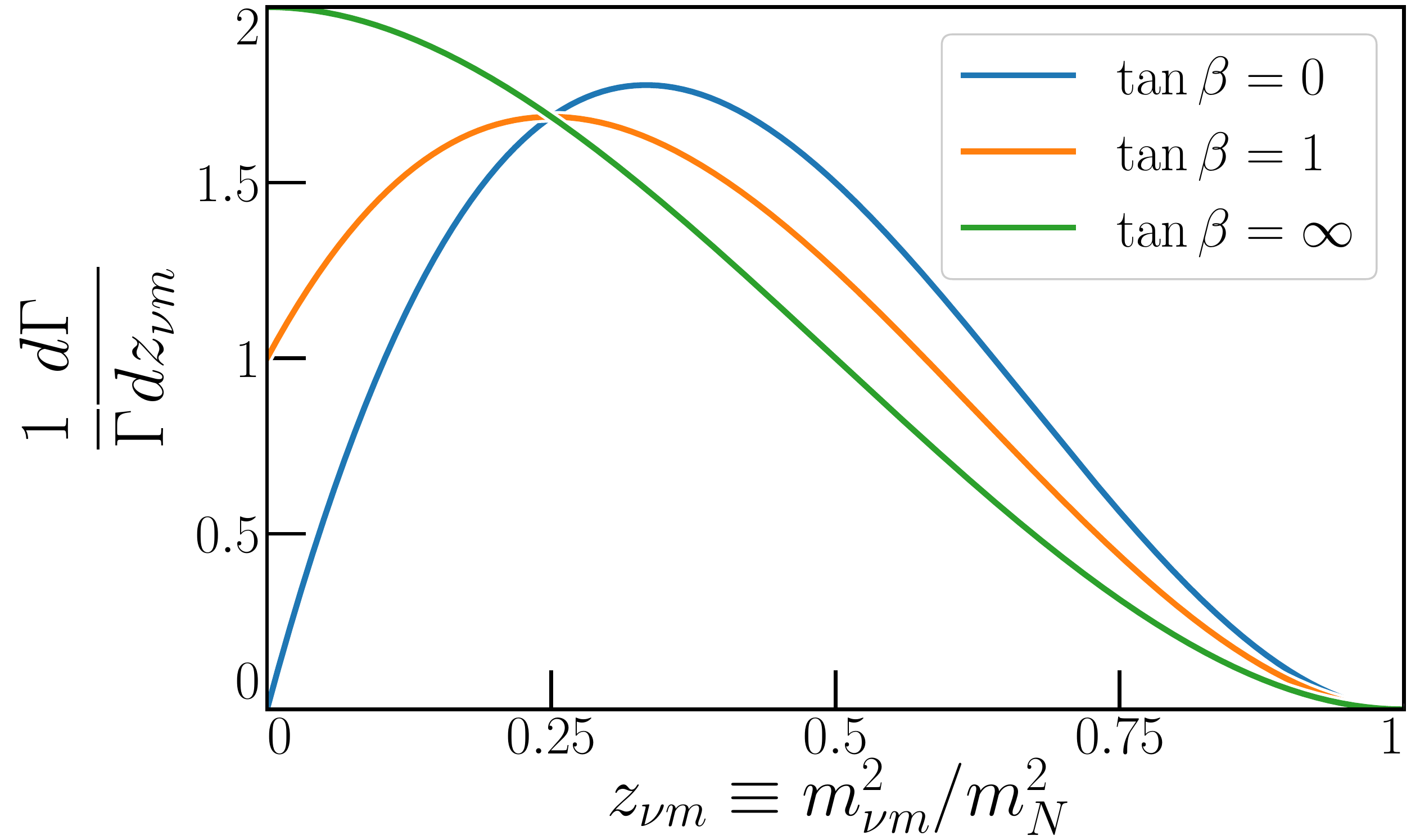}
\caption{One-dimensional decay distributions as a a function of the (reduced) neutrino/negatively-charged-lepton invariant mass for vector/axial-vector interactions assuming three different choices of $\tan\beta$ as labelled -- see Eq.~\eqref{eq:CsNoPolMaj}. We assume that $m_N \gg m_e$ in generating these distributions.
\label{fig:TanBznum}}
\end{center}
\end{figure}

As discussed throughout Section~\ref{subsec:MFDF:NoSpin}, for a DF with vector/axial-vector interactions, the decay distributions can vary significantly depending on $\tan\beta$. Differentiating between different $\tan\beta$ can be difficult, exemplified by the degeneracies present in Fig.~\ref{fig:HNLUSq}(left). In practice, one observable that is useful in differentiating between different $\tan\beta$ under the DF vector/axial-vector hypothesis is the one-dimensional distribution as a function of $z_{\nu m}$. This distribution can be expressed as
\begin{equation}
\dfrac{1}{\Gamma}\dfrac{d\Gamma}{dz_{\nu m}} = \frac{2\left(1 - z_{\nu m}\right)^2}{1 + \tan^2\beta} \left( \tan^2\beta + 2\left(3 + \tan^2\beta\right)z_{\nu m}\right),
\end{equation}
which has very different behavior for different choices of $\tan^2\beta$. We display this distribution for $\tan\beta = 0,\ 1,\ \infty$ in Fig.~\ref{fig:TanBznum}.

To numerically estimate the required number of events to separate these different hypotheses significantly, we perform a similar analysis to that of Section~\ref{subsec:MFDF:YesSpin} where we simulate pseudoexperiments assuming some truth hypothesis and test two different hypotheses -- a null hypothesis and an alternative one. As a reminder, in this unbinned analysis, we use the full five-dimensional phase space of the $N \to \nu e^+ e^-$ decay to extract maximal information, which includes the two- and one-dimensional discussion surrounding Figs.~\ref{fig:SVSpinInformation} and~\ref{fig:TanBznum}. We determine, given those pseudoexperiments and the two hypotheses' fits to the pseudodata, the expected number of events required to prefer the null hypothesis over the alternative one at high confidence. The three truth hypotheses that we explore are Cases 1-3, defined in Section~\ref{subsec:MFDF:YesSpin}.

\begin{figure}
\begin{center}
\includegraphics[width=0.7\linewidth]{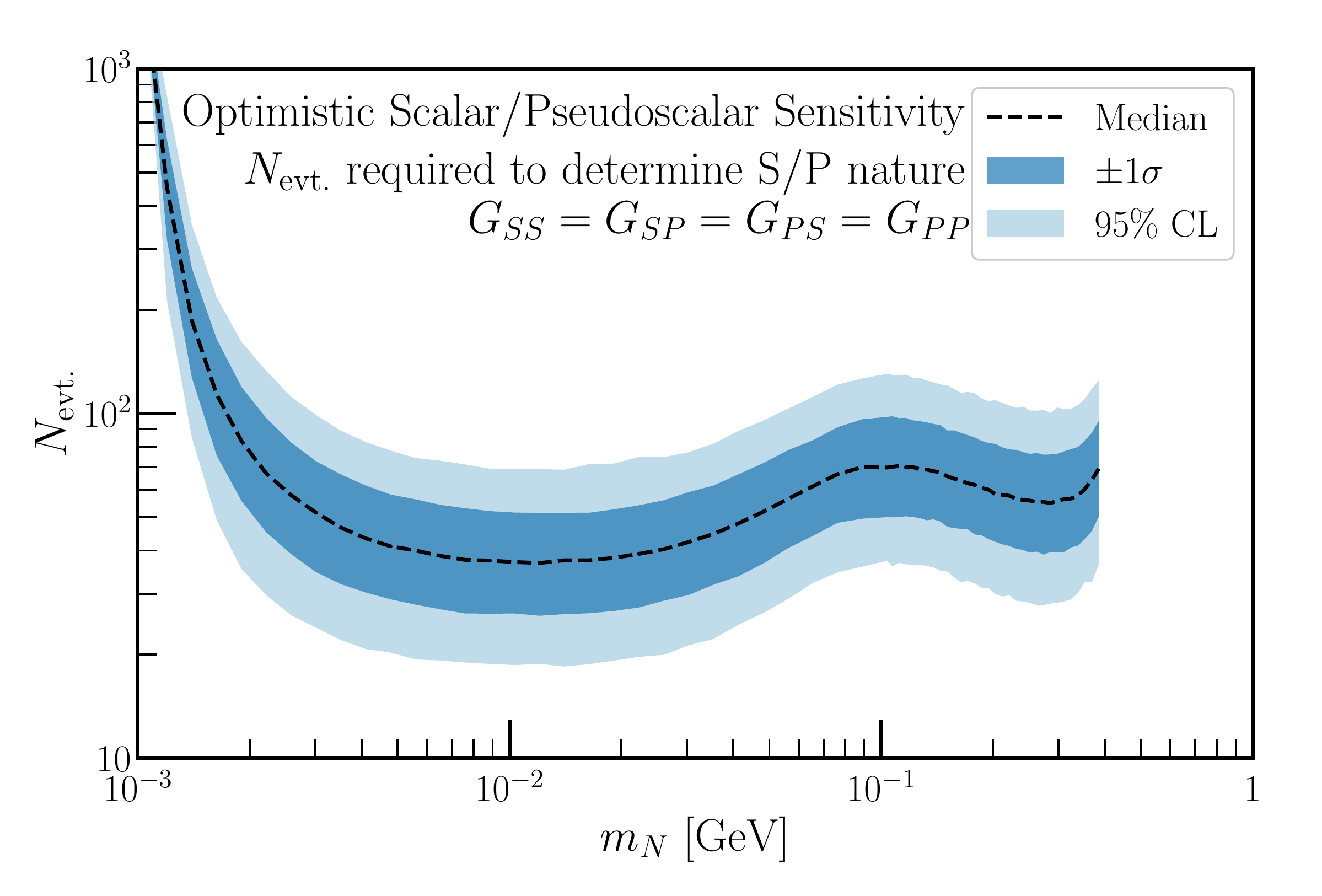}
\caption{Number of events as a function of $m_N$ required to prefer scalar/pseudoscalar interactions over vector/axial-vector/tensor ones at $3\sigma$ assuming that the Dirac Fermion hypothesis, with $G_{SS} = G_{SP} = G_{PS} = G_{PP}$, is true. See Appendix~\ref{app:Unbinned} for detail. The dashed black line corresponds to the median expected capability, whereas the dark (light) blue filled regions show the $\pm 1\sigma$ ($95\%$) expectation.
\label{fig:IntStr:YesSpin:SP}}
\end{center}
\end{figure}

Fig.~\ref{fig:IntStr:YesSpin:SP} displays the results of this procedure when the simulated data are consistent with only scalar or pseudoscalar couplings, with the structure as labelled on the figure. In fitting these pseudodata, we compare two hypotheses: the null hypothesis is that $N$ is a DF with any possible couplings $G_{NL}$ allowed to be nonzero, and the alternative hypothesis still assumes $N$ is a DF, but fixes $G_{SS} = G_{SP} = G_{PS} = G_{PP} = 0$. In determining the preference for the null hypothesis over the alternative one, we are determining the preference for scalar/pseudoscalar interactions, over vector/axial-vector/tensor ones.
In the analysis that leads to Fig.~\ref{fig:IntStr:YesSpin:SP}, we have assumed that the nature of $N$, i.e. that it is a DF, is known. We see that, for this choice of true couplings, $G_{SS} = G_{SP} = G_{PS} = G_{PP}$, it is in fact easier to determine the interaction structure (scalar/pseudoscalar vs. vector/axial-vector/tensor) than it is to determine the nature (DF vs. MF, cf Fig.~\ref{fig:MFDF:YesSpin:SP}). We also note that, when $m_N \approx m_\mu$ and $N$ is unpolarized, there is still the possibility of determining the interaction structure. We will focus on this scenario in Section~\ref{subsec:IntStr:NoSpin}.

\begin{figure}
\begin{center}
\includegraphics[width=\linewidth]{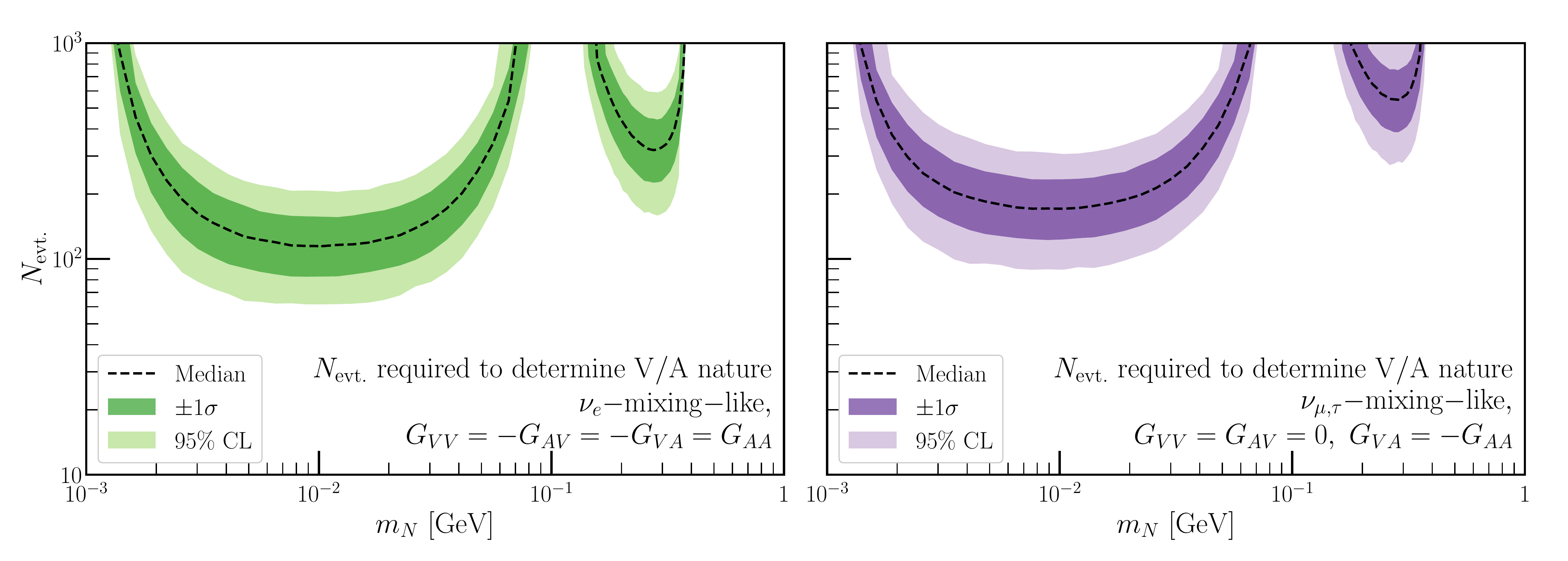}
\caption{Number of events as a function of $m_N$ required to prefer vector/axial-vector interactions over scalar/pseudoscalar/tensor ones at $3\sigma$ assuming that the Dirac Fermion hypothesis is true. Two different assumptions are made about the true distributions, as labelled on the respective figures. See Appendix~\ref{app:Unbinned} for more detail. The dashed black line corresponds to the median expected capability, whereas the dark (light) filled regions show the $\pm 1\sigma$ ($\pm 95\%$) expectation.
\label{fig:IntStr:YesSpin:VA}}
\end{center}
\end{figure}

We compare the other data that we simulate -- consistent with $N$ being a DF with vector/axial-vector interactions of either $\nu_{e}$-like or $\nu_{\mu}$/$\nu_{\tau}$-like mixing scenarios -- against two hypotheses as well. The null hypothesis is the same as above, that $N$ is a DF and all $G_{NL}$ are allowed. The alternative hypothesis is that $G_{VV} = G_{VA} = G_{AV} = G_{AA} = 0$, and the comparison then determines the preference for vector/axial-vector interactions over scalar/pseudoscalar/tensor ones. The results of this are shown in Fig.~\ref{fig:IntStr:YesSpin:VA}, with the $\nu_e$- ($\nu_{\mu,\tau}$-)mixing scenario shown in the left (right) panel.
Both simulated data assumptions have similar results. For $m_{N} \approx 2m_{e}$, determining that $N$ decays via vector/axial-vector interactions is difficult. This is due to the fact that, when $m_e/m_N$ is non-negligible, many contributions from different $G_{NL}$ contribute to the matrix-element-squared (see Ref.~\cite{deGouvea:2021ual}), meaning that scalar/pseudoscalar/tensor interactions may adequately mimic the vector/axial-vector ones. Similarly, differentiating between vector/axial-vector and scalar/pseudoscalar/tensor scenarios is difficult for both panels of Fig.~\ref{fig:IntStr:YesSpin:VA} when $m_N \approx m_\mu$ and the $N$ source is unpolarized. As we will demonstrate in Section~\ref{subsec:IntStr:NoSpin}, this is because, when $N$ is unpolarized and $m_N \gg m_e$, the combination of scalar/pseudoscalar/tensor interactions can nearly perfectly mimic the vector/axial-vector ones. Lastly, we briefly comment on the small differences between the left and right panels of Fig.~\ref{fig:IntStr:YesSpin:VA}. Generically, for a given $m_N$, we see that slightly fewer events are required to establish the vector/axial-vector preference for the $\nu_e$-mixing-like (green, left panel) scenario than for the $\nu_{\mu,\tau}$-mixing-like (purple, right panel) scenario. This is because the $\nu_e$-mixing like scenario results in small $\tan\beta$ (see Eqs.~\eqref{eq:CsNoPolDir},\eqref{eq:CsNoPolMaj}), whereas the $\nu_{\mu,\tau}$-mixing scenario results in $\tan\beta \approx 1$. Scalar/pseudoscalar/tensor interactions can mimic vector/axial-vector ones more efficiently when $\tan\beta = 1$ than when $\tan\beta = 0$.

In comparing Figs.~\ref{fig:MFDF:YesSpin:SP}-\ref{fig:MFDF:YesSpin:VA} with Figs.~\ref{fig:IntStr:YesSpin:SP}-\ref{fig:IntStr:YesSpin:VA}, we see that comparable event rates are required to distinguish MF from DF and to determine the coupling structure once the nature is known. Generically, model determinations may be obtained with $\mathcal{O}(50)$ events if $N$ is a DF decaying via scalar/pseudoscalar interactions (with favorable couplings) and $\mathcal{O}(100)$ if its decays are via vector/axial-vector interactions. These numbers should serve as a benchmark for planning an experiment in the wake of the discovery of a new particle $N$, hoping to explore its properties in greater detail.

%%%%%%%%%%%%%%%%%%%%%%%%%%%%%%%%%%%%%%%%%%%%
\subsection{Observations with an Unpolarized Source}\label{subsec:IntStr:NoSpin}\setcounter{footnote}{0}
%%%%%%%%%%%%%%%%%%%%%%%%%%%%%%%%%%%%%%%%%%%%
%In Section~\ref{subsec:IntStr:YesSpin}, we noted that several features occur when hoping to determine the interaction structure of $N$ and when $m_N \approx m_\mu$, resulting in an unpolarized source. 
Section~\ref{subsec:MFDF:NoSpin}, focused on how, with an unpolarized $N$ source, we can still determine whether $N$ is a DF or a MF. Here, we return to that framework, focusing instead on determining the structure of the interaction that mediates $N$ decay.
%If $N$ is discovered (and identified as a DF or MF) through two-body decays, then its interaction structure should be trivial to determine. With the general three-body decay framework we have introduced, it is less straightforward if $N$ is discovered via three-body decays of the type $N \to \nu \ell_\alpha^- \ell_\alpha^+$.
As we did in Section~\ref{subsec:MFDF:NoSpin}, we make some simplifying assumptions. We focus on the $\alpha = \beta$ scenario, in which the outgoing final-state charged leptons are identical. We also focus on the limit where $m_{\ell}/m_N \to 0$, and can apply the expressions obtained in Eqs.~\eqref{eq:CsNoPolDir},\eqref{eq:CsNoPolMaj} for this analysis. We perform a demonstrative case study, where we assume that $N$ is known to be a MF\footnote{This can be deduced, for example, via observations of other properties regarding $N$ or the light neutrinos, such as the observation of neutrinoless double beta decay -- if the light, SM neutrinos are MF then the $N$ is most likely also a MF since, by assumption, it mixes with the active neutrinos.}, and explore how well we can determine that its decays are mediated by scalar/pseudoscalar, vector/axial-vector, or tensor interactions.

\begin{figure}
\begin{center}
\includegraphics[width=\linewidth]{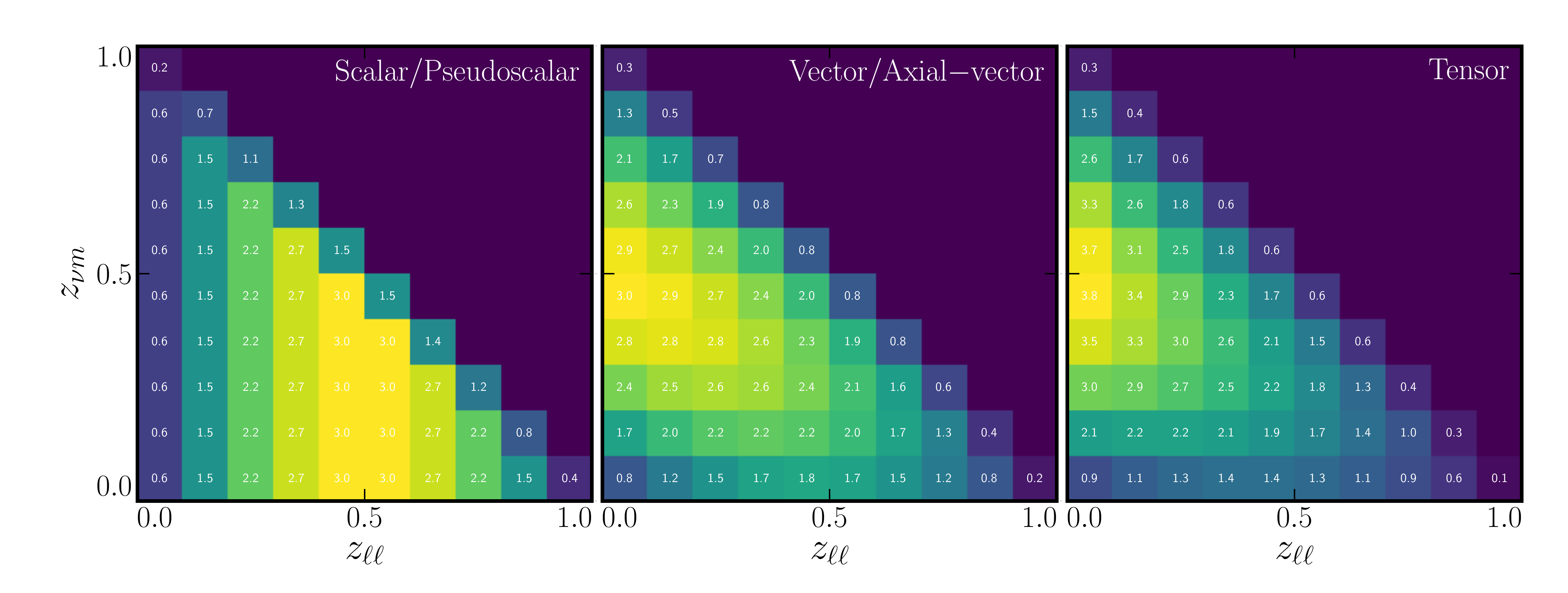}
\caption{Expected event distributions as a function of $z_{\ell\ell} = m_{\ell\ell}^2/m_N^2$ and $z_{\nu m} = m_{\nu m}^2/m_N^2$ assuming that $N$ is a Majorana fermion with decays mediated by scalar/pseudoscalar interactions only (left), vector/axial-vector interactions only (center), or tensor interactions only (right). Parameters are chosen such that the total event rates (given by the labels in each bin, yellow corresponding to more events) are $100$ in each case.
\label{fig:SVT:EvtDist}}
\end{center}
\end{figure}

We simulate three sets of pseudodata as a function of $z_{\ell\ell}$ and $z_{\nu m}$, assuming that $100$ signal events are expected for different truth combinations of $S$, $V$, and $T$. Fig.~\ref{fig:SVT:EvtDist} displays the expected distribution of these signal events according to the three different sets we generate -- the left panel corresponds to scalar/pseudoscalar contributions only ($S = 1$, $V = T = 0$), the center corresponds to vector/axial-vector only ($V = 1$, $S = T = 0$), and the right panel corresponds to tensor contributions ($T = 1$, $S = V = 0$). The distinct shapes of the distributions across $z_{\ell\ell}$ and $z_{\nu m}$ allow for differentiation between different contributions (including, potentially, multiple structures contributing).

\begin{figure}
\begin{center}
\includegraphics[width=0.7\linewidth]{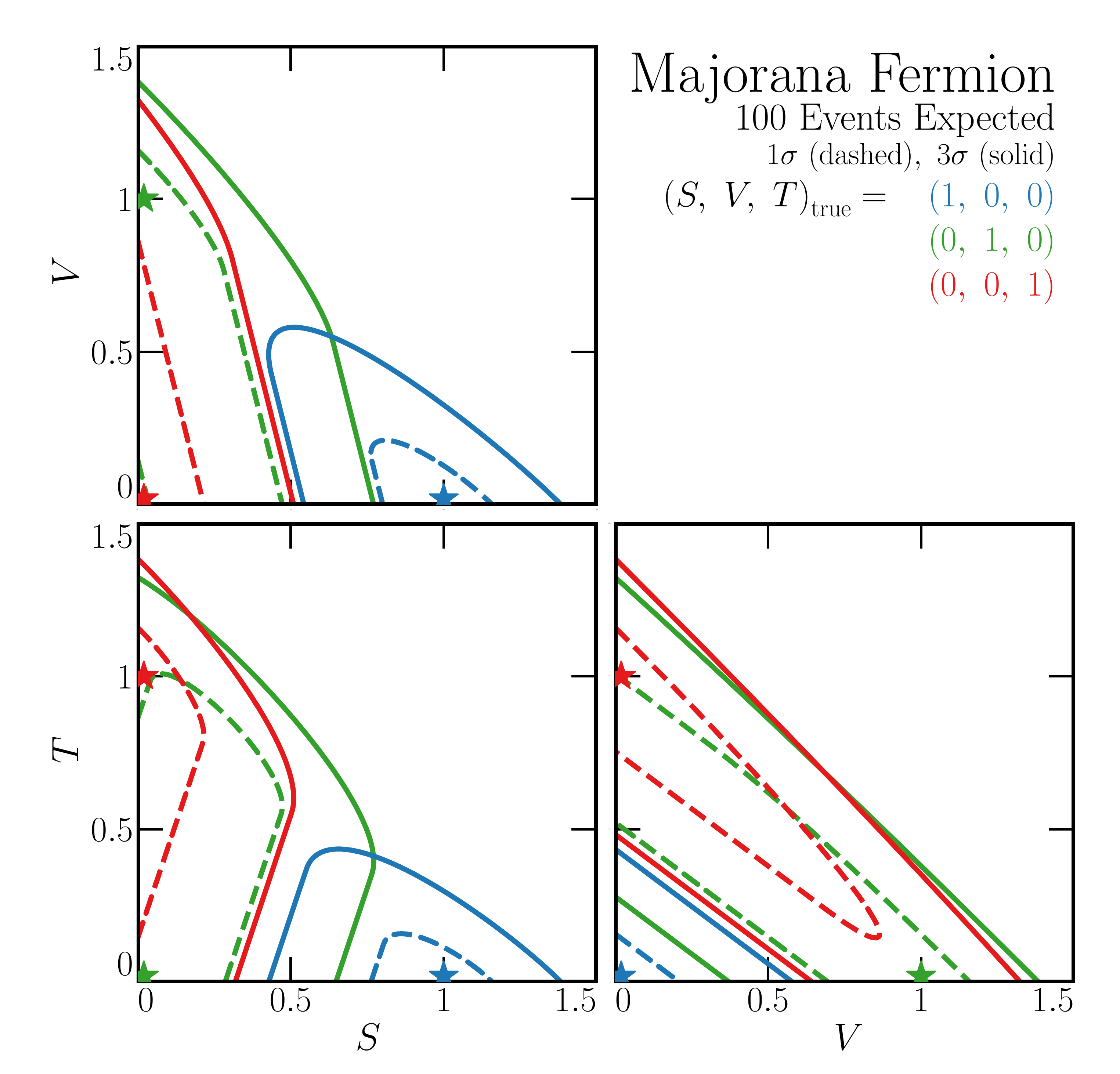}
\caption{Measurement capability for the parameters $\left(S,\ V,\ T\right)$ assuming $N$ is a Majorana fermion and 100 signal events are expected. The three colors indicate fits to data simulated with three different assumed true combinations of $\left(S,\ V,\ T\right)$ as indicated in the legend. Dashed (solid) lines correspond to $1\sigma$ ($3\sigma$) confidence level measurements, and stars indicate the best-fit point of each fit.
\label{fig:MajMeasSVT}}
\end{center}
\end{figure}

For each of these three datasets, we perform a fit to $\left(S,\ V,\ T\right)$ to determine the capability of measuring these parameters if $100$ signal events are expected. The results of these three fits are shown in Fig.~\ref{fig:MajMeasSVT} -- blue contours correspond to measuring the scalar/pseudoscalar dataset, green contours correspond to the vector/axial-vector one, and red contours correspond to the tensor case. For all three datasets, we present $1\sigma$ and $3\sigma$ measurements of the parameters as dashed and solid lines, respectively.
Each panel presents the two-dimensional (marginalized over the unseen, third parameter) measurement of two of the three parameters $\left(S,\ V,\ T\right)$, and colored stars in each panel indicate the best-fit point of each of the separate fits.

A few features of Fig.~\ref{fig:MajMeasSVT} are noteworthy -- If the decays of $N$ are purely scalar/pseudoscalar (blue contours), the hypothesis $S = 0$ can be excluded at high significance. Likewise, both $V$ and $T$ are constrained to be significantly smaller than $1$ (bottom right panel). These features arise because if $S$ is the only nonzero parameter, our expected $\left(C_4,\ C_5,\ C_6\right) = \left(0,\ 0,\ 1\right)$, an arrangement that cannot be mimicked by any combination of nonzero $V$ and $T$. On the other hand, if $V$ is the only nonzero parameter (green contours), the hypothesis $V = 0$ can be accommodated easily with the combination $T = 0.75$, $S = 0.25$. In fact, these two combinations $(0, 1, 0)$ and $(0.25, 0, 0.75)$ are indistinguishable without polarization. Finally, if the decays of $N$ are purely of tensor structure, $T = 0$ can be excluded at low (${\sim}1\sigma$) confidence -- nonzero $S$ and $V$ can almost, but not perfectly, mimic this case.

Whether we can actually determine that $N$ is a MF and perform this analysis is not guaranteed. However, we see this as a useful exercise given that the differential decay distribution depends only on three model-related parameters (in contrast to the five free parameters of the DF hypothesis in Eq.~\eqref{eq:CsNoPolDir}). Such fits may be performed under the DF hypothesis with this extra freedom. Given the structure of $\left( C_4,\ C_5,\ C_6\right)$ under the DF hypothesis in Eq.~\eqref{eq:CsNoPolDir}, we can conclude that the easiest structures to identify are the pure-scalar one (only $S$ is nonzero), and ones in which $V$ is the only nonzero parameter with $\tan^2\beta \to 0$ or $\infty$. These three scenarios predict that $\left( C_4,\ C_5,\ C_6\right) = \left(1,\ 0,\ 0\right)$, $\left(0,\ 1,\ 0\right)$, or $\left(0,\ 0,\ 1\right)$, which are easiest to properly identify.

%%%%%%%%%%%%%%%%%%%%%%%%%%%%%%%%%%%%%%%%%%%%
\section{Discussion \& Conclusions}
\label{sec:Conclusions}

Many currently-operating experiments, as well as those in the planning stages, intend to carry out searches for a wide array of beyond-the-Standard-Model physics.  In this work, we have focussed on the production of light and relatively long-lived particles produced in meson decays which can transit from a beam dump to a large volume detector where they subsequently interact or decay. 
%
%Given this, it is prudent to plan for such a potential discovery. More specifically, many of these experiments seek to discover a new particle decaying into SM particles. 
%
Many classes of models (dark photons, dark Higgs bosons, heavy neutral leptons, etc.) fit this description, and most experimental efforts are capable of probing these different model classes simultaneously. In the event of a discovery, characterizing the newly-discovered particle(s) will be a high priority.

We have focused on the scenario in which a heavy fermion is discovered and that we wish to determine several of its properties, concentrating on two specific issues: (a) whether the fermion is of the Dirac type (where Lepton Number is a conserved quantity, including of the new particles) or whether it is of the Majorana type and is its own antiparticle, and (b) whether the new particle's decays are mediated by additional force-carrying particles beyond those in the SM. 

In order to address this possible future scenario, we have envisioned a ``post-discovery'' experiment that is purpose-built to study the new particles' properties. We have focused on the MeV to GeV mass range for new particles and have restricted ourselves to the scenario in which the new particle can be produced via meson decay-at-rest. A suitable experimental approach then (schematized in Fig.~\ref{fig:ExpSetup}) is to have a proton source impinge on a target with a detector nearby in a direction perpendicular to the beam direction in order to reduce beam-induced backgrounds. The detector is imagined to be a low-density gaseous argon time-projection chamber embedded in a magnetic field, allowing for precision reconstruction of the new particle's decay products, especially electrons. The polarization of any new particle being produced in this process is fundamental to understanding its properties, an aspect we explored in some detail.

To answer the question of DF versus MF, we have built on our previous analysis~\cite{deGouvea:2021ual} of the details of the decay distributions of fermions into two-body and three-body final states, which exhibit distinct properties under these two separate hypotheses. Moving beyond this previous analysis, we have quantified the required number of observed events in an experiment that is necessary to definitively determine the new fermion's DF/MF nature, taking advantage of the fully differential decay distribution.  While our analysis technique can be applied to arbitrary couplings of the new light state, we have focused on a few well motivated examples and quantified in each case the number of signal events necessary to distinguish a DF with specific couplings from a MF with general couplings.  We have found that, in some generic new-physics model for the decay of the heavy fermion, hundreds of events will be required for this separation. With some favorable model-parameters, this separation could be performed with as few as three dozen events.  Furthermore, in some cases this separation is robust under a loss of polarization in the production of the new fermion while for others if $P=0$ the two hypotheses cannot be distinguished.  In an actual experiment the polarization is expected to be non-zero, except for the situation where the new fermion is mass degenerate with a SM lepton.

Determining what type of interaction (i.e., scalar, pseudoscalar, vector, axial-vector, tensor, or some combination thereof) mediates the new particle's decays would also be of interest since the answer can help determine whether new force carriers beyond those of the SM exist. We have explored how well the coupling structure could be determined if the DF/MF nature of the fermion is known.  
Similar to the MF/DF separation question, we have found that hundreds of events are required to perform reasonable separation between different coupling hypotheses.  Intriguingly, a scalar/pseudoscalar-coupled DF can be distinguished from a DF without scalar couplings more easily than it can be determined to be DF.  As before, in some cases (not the same cases as above) the ability to identify the correct coupling structure is robust even if polarization is lost.  For the particular $P=0$ case of a MF, only the scalar-coupled fermion can be well distinguished from vector/tensor-coupled but the converse is not true.

As current/next-generation experiments begin their searches for new particles, it is crucial to be prepared for any outcome so as to best carry forward this experimental program. Beyond the analyses we have carried out in this work, one can imagine similar studies for other classes of new-physics particles (scalars, vector bosons, etc.) and other signatures beyond particle decays. The possible discovery of not only a new particle decaying, but evidence of further new-physics particles mediating its decays, is tantalizing. Regardless, a roadmap for what comes after the field's next discoveries is warranted in this very exciting time in particle physics phenomenology.

%%%%%%%%%%%%%%%%%%%%%%%%%%%%%%%%%%%%%%%%%%%%

\section*{Acknowledgements}
The work of AdG is supported in part by the DOE Office of Science
award \#DE-SC0010143.
PJF, BJK, and KJK are supported by Fermi Research Alliance, LLC under Contract DE-AC02-07CH11359 with the U.S. Department of Energy. AdG and KJK thank the Institute for Nuclear Theory at the University of Washington for its hospitality during which a portion of this work was completed. This work was performed in part at the Aspen Center for Physics, which is supported by the National Science Foundation grant PHY-1607611.

\appendix
%\clearpage
%\newpage

\section{Unbinned Likelihood Analyses}\label{app:Unbinned}

The Neyman-Pearson lemma states that the most powerful way to differentiate between two hypotheses is to use their likelihood ratio.  In order to extract as much information out of a set of measurements as possible, we calculate likelihoods for unbinned data. In Section~\ref{sec:DirVsMaj} we attempt to differentiate between the hypotheses that $N$ is either a DF or a MF; in Section~\ref{sec:InteractionStructure} we assume that the DF/MF nature is known and attempt to differentiate between the hypotheses that its decays are mediated by scalar/pseudoscalar or vector/axial-vector interactions. In this appendix, we explain the procedure we perform for these unbinned likelihood analyses and the metrics we adopt to deem one hypothesis preferred over another.

As a proxy for real experimental data we will perform pseudoexperiments where we generate mock data under specific assumptions for the underlying physics, e.g., the HNL is a DF coupled to the SM only through mixing with $\nu_\mu$, and for the expected number of HNLs produced.  We then attempt to fit these data with alternative hypotheses, and determine for each case what choice of coupling parameters best fits the data by maximizing the (log) likelihood.  To compare the hypotheses we use the likelihood ratio test.  In the limit of large statistics the likelihood ratio is expected to follow a $\chi^2$-distribution.  However, we will often be considering the opposite limit and will determine the discriminatory power through Monte Carlo i.e., we carry out the generation and fitting of data many times and study the distribution of results.  

For concreteness, consider trying to distinguish between the DF and MF hypotheses. The likelihood ratio is defined as
\begin{equation}
\lambda \equiv \frac{\underset{\theta_{\rm DF}}{\mathrm{max}}\left\lbrace \mathcal{L}(\{\vec{x}\};\theta_{\rm DF}\right\rbrace}{\underset{\theta_{\rm MF}}{\mathrm{max}}\left\lbrace \mathcal{L}(\{\vec{x}\};\theta_{\rm MF}\right\rbrace},
\end{equation}
where $\mathcal{L}(\{\vec{x}\}; \theta_{\rm H})$ is the likelihood function for a hypothesis H given set of data $\{\vec{x}\}$.  Each datum is a vector $\vec{x}$ in the $5$ dimensional phase space.
For convenience we use the extended likelihood approach \cite{Barlow:1990vc}, and the likelihood function is 
\begin{equation}
\mathcal{L}(\{\vec{x}\}, \theta_{\rm H}) \equiv \frac{e^{-\mu_{\rm H}} \mu_{\rm H}^n}{n!} \prod_{i=1}^{n}\lvert \mathcal{M}(\vec{x}_i, \theta_{\rm H})\rvert^2~,
\end{equation}
with $\mathcal{M}$ given in, for example, Eq.~(\ref{eq:MGNL}).  This likelihood is maximized when the number of expected events, $\mu_{\rm H}$, is the same as the observed number, $n$.  The number of expected events is determined by the integral over all of phase space of the partial width, Eq.~(\ref{eq:diffxsec}).  We choose to normalize the $G_{NL}$ such that the largest of these parameters in the fit is $1$ and the overall normalization, which determines $\mu_{\rm H}$, will adjust to accommodate the observed number of events, which varies for each pseudoexperiment.  Thus, we only use the relative sizes of couplings to distinguish between hypotheses.

Our goal then is to perform a large number of pseudoexperiments and determine the distribution of $\lambda$ and determine whether, for a given truth assumption (couplings and number of expected signal events), the alternative hypothesis can be robustly excluded. In order to determine what threshold value of $\lambda$ corresponds to a significant preference, we must account for the number of free parameters associated with each hypothesis $\theta_{\rm H}$.
When attempting to distinguish between the DF and MF hypotheses in Section~\ref{sec:DirVsMaj}, the DF hypothesis contains 18 free parameters while the MF hypothesis contains 9. The MF hypothesis is a nested subset of the DF one, and can be realized under certain restrictions on the DF free parameters~\cite{deGouvea:2021ual}, so excluding the alternative (MF) hypothesis can be viewed as performing an exclusion over 9 free parameters. We are interested in $3\sigma$ exclusion, so for nine free parameters, $\Delta \chi^2 = 25.26$ is appropriate. Given the relative factor of two between $\Delta \chi^2$ and $-\log\lambda$, we use $-\log\lambda = 12.63$ as our critical threshold. In Section~\ref{sec:InteractionStructure}, we perform comparisons involving hypotheses with 18 and 10 free parameters -- the difference of 8 parameters causes us to use $-\log\lambda = 11.79$ as our $3\sigma$ criterion.

\begin{figure}
\centering
\includegraphics[width=\linewidth]{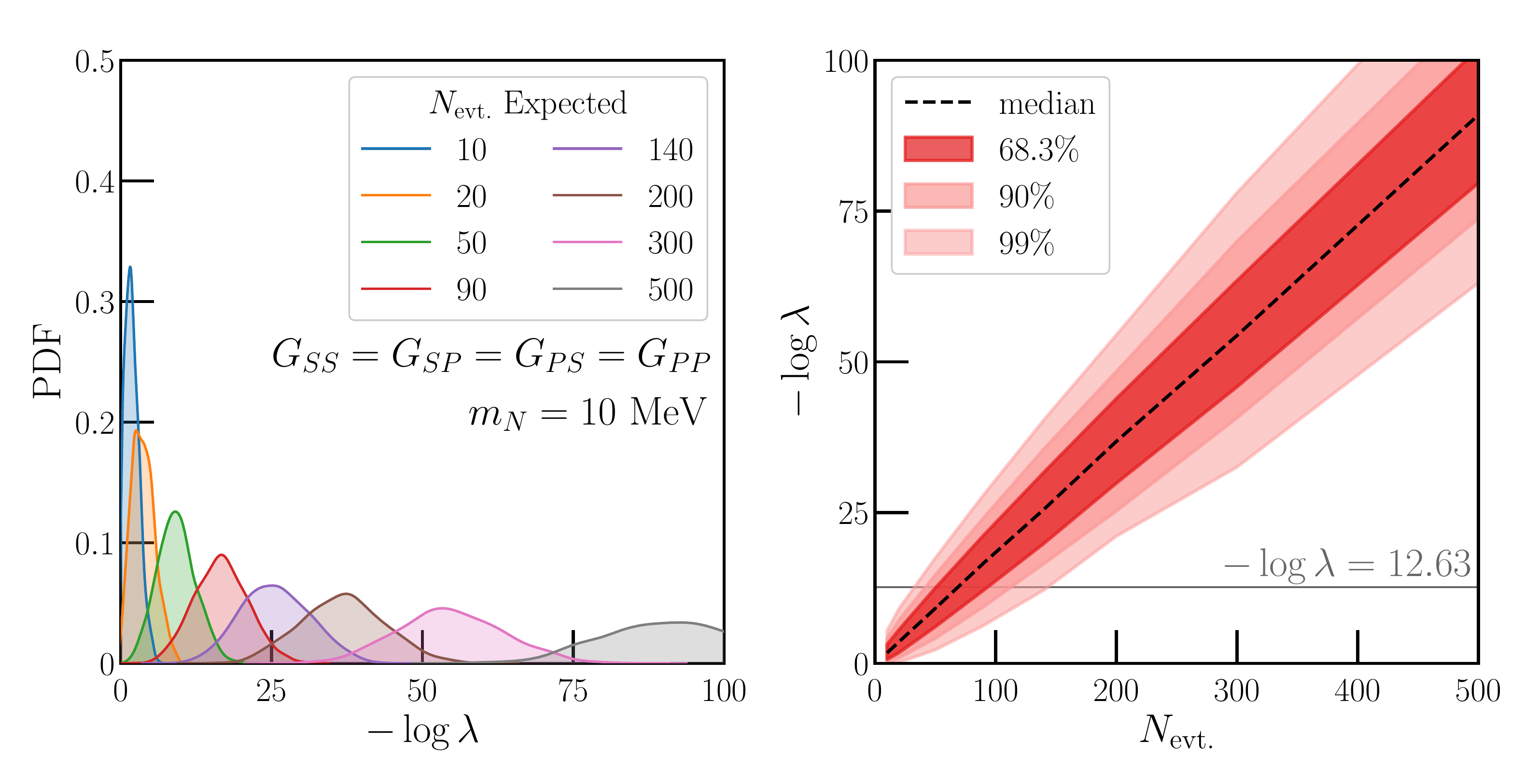}
\caption{Expected distribution of the difference of negative-log-likelihoods, $-\log \lambda$, between the DF and MF hypotheses after performing pseudoexperiments with the truth assumed to be a DF. Eight different distributions are shown for $n_0 \in \left\lbrace 10, 20, 50, 90, 140, 200, 300, 500\right\rbrace$, with 2000 pseudoexperiments each.
\label{fig:DeltaLambda}}
\end{figure}

As a concrete example of the situation where we attempt to distinguish between the DF and MF hypotheses, we consider a case where $N$ is a DF with mass $m_N = 10$ MeV and has only scalar and pseudoscalar couplings arranged in a way that the decay distribution's forward/backward asymmetry is maximized. We then perform pseudoexperiments, as described above, for a number of different expected signal events $n_0 \in \left\lbrace 10, 20, 50, 90, 140, 200, 300, 500\right\rbrace$. The resulting distributions of $-\log\lambda$ for these eight different cases are shown in the left panel of Fig.~\ref{fig:DeltaLambda}.
The relevant quantity we desire in this case is the median of the $-\log\lambda$ distribution for each value of $n_0$. We extract the median, as well as the $68.3\%$, $90\%$, and $99\%$ ranges of these distributions as a function of $n_0$ and display that in the right panel of Fig.~\ref{fig:DeltaLambda}. Since our goal is to determine the number of expected events required to distinguish between the two hypotheses at a high enough confidence, we determine where the dashed line (median) intersects $-\log\lambda = 12.63$. In this case, for the scalar/pseudoscalar couplings and $m_N = 10$ MeV, we find that this occurs for $n_0 \approx 70$.

\bibliographystyle{JHEP}
\bibliography{refs}

\end{document}